\documentclass[12pt,preprint]{aastex}

\def\fuse{{\it FUSE\/}}

\def\hmol{H$_{\rm 2}$\/}

\def\kms{km s$^{-1}$}

\def\mfarcs{\hbox{$~\!\!^{\prime\prime}$}}

\slugcomment{Version \today}
\shorttitle{D/H toward 2 stars}
\shortauthors{Oliveira et al.}

\begin{document}

\title{Variations in the D/H ratio of extended sightlines from {\it FUSE} observations\altaffilmark{1}}

\author{Cristina M. Oliveira\altaffilmark{2} and Guillaume H{\'e}brard\altaffilmark{3}}
  
\altaffiltext{1}{Based on observations made with the NASA-CNES-CSA {\it Far Ultraviolet Spectroscopic Explorer}. \fuse~is operated for NASA by The Johns Hopkins University under NASA contract NAS5-32985.}
\altaffiltext{2}{Department of Physics and Astronomy, The Johns Hopkins University, 3400 N. Charles St., Baltimore MD 21218}
\altaffiltext{3}{Institut d'Astrophysique de Paris, 98 bis, Boulevard Arago, 75014 Paris, France.}

\begin{abstract} 

We use new {\it FUSE} data to determine the column densities of interstellar D\,I, N\,I, O\,I, Fe\,II, and \hmol~along the HD\,41161 and HD\,53975 sightlines. Together with $N$(H\,I) from the literature (derived from {\it Copernicus} and {\it IUE} data) we derive D/H, N/H, and O/H ratios. These high column density sightlines have both log $N$(H\,I) $>$ 21.00 and allow us to probe gas up to $\sim$1300 pc. In particular these sightlines allow us to determine the gas phase D/H ratio in a hydrogen column density range, log $N$(H) $>$ 20.70, where the only five measurements available in the literature yield a weighted average of D/H = (0.86 $\pm$ 0.08)$\times10^{-5}$. We find D/H = (2.14 $\pm~^{0.51}_{0.43}$)$\times10^{-5}$ along the HD\,41161 sightline. This ratio is $\sim$3$\sigma$ higher than the weighted mean D/H ratio quoted above, for sightlines with log $N$(H) $>$ 20.70, while the D/H ratio for the HD\,53975 line of sight, D/H = (1.02 $\pm~^{0.23}_{0.20}$)$\times10^{-5}$, agrees within the 1$\sigma$ uncertainties. Our D/H measurement along the HD\,41161 sightline presents the first evidence of variations of D/H at high $N$(H). Our result seems to indicate that either the long sightlines that according to the deuterium depletion model are dominated by cold undisturbed gas where deuterium would be depleted onto carbonaceous grains occur at higher $N$(H) than previously thought or that the clumping of low D/H values in the literature for the long sightlines has another explanation. Another possibility to explain the high D/H toward HD\,41161 is local production of D\,I which cannot be ruled out. Both of the O/H ratios derived here, (9.12 $\pm~^{2.15}_{1.83}$)$\times10^{-4}$ and (5.37 $\pm~^{1.35}_{1.14}$)$\times10^{-4}$ (for HD\,41161 and HD\,53975, respectively) are higher than what has been found by other authors in studies containing many sightlines, although not inconsistent, in the case of HD\,53975, with individual measurements along some sightlines. However, O/H along HD\,41161 is extremely high, and together with other high O/H ratios based on {\it FUSE} data might indicate the presence of unknown systematic effects when deriving $N$(O\,I) along high column density sightlines. Finally, we derive (N/H)$\times10^{5}$ =  8.32 $\pm~^{2.09}_{1.76}$ and 5.07 $\pm~^{1.45}_{1.21}$ and (D/O)$\times10^2$ = 2.29 $\pm~^{0.40}_{0.35}$ and 1.91 $\pm~^{0.51}_{0.43}$, for HD\,41161 and HD\,53975, respectively. In addition, the relatively high signal-to-noise ratio of the HD\,41161 data allows us to place constraints on the $f$-values of some neutral chlorine transitions, present in the {\it FUSE} bandpass, for which only theoretical values are available.

\end{abstract}


\section{INTRODUCTION} 
\label{intro}

Deuterium is thought to be produced in appreciable amounts only in the Big Bang \citep{1976Natur.263..198E}. Of all the light elements produced during Big Bang Nucleosynthesis (e.g., $^3$He, $^4$He, $^7$Li), deuterium, easily destroyed in stellar interiors (astration), is the one whose relatively simple evolution makes it a particularly sensitive estimator of the baryonic density. Conversely, measurements of the cosmic microwave background, such as those performed with {\it WMAP} and other missions, can be used to infer the primordial abundance of deuterium, (D/H)$_{\rm P}$. Analysis of the three year {\it WMAP} data \citep{2006astro.ph..3449S} yields (D/H)$_{\rm P}$ in agreement with that derived by \citet{2003ApJS..149....1K}, (D/H)$_{\rm P}$ = (2.78 $\pm~^{0.44}_{0.38}$)$\times10^{-5}$, by taking the weighted average of five D/H measurements toward QSOs in the range $z$ = 2.079--3.572. \citet{2004ApJS..150..387S} found D/H = (2.2 $\pm$ 0.7)$\times10^{-5}$ for Complex C, a high-velocity low-metallicity cloud falling into our galaxy, which has presumably experienced more stellar processing than the gas seen toward QSOs.

Precise measurements of the gas-phase D/H ratio in the ISM were first performed with the {\it Copernicus} satellite \citep[e.g.,][]{1973ApJ...186L..95R} followed by the {\it Hubble Space Telescope} twenty years later \citep[{\it HST}, e.g.,][]{1995ApJ...451..335L}, IMAPS \citep[Interstellar Medium Absorption Profile Spectrograph,][]{1999ApJ...520..182J,2000ApJ...545..277S}, and more recently the {\it Far Ultraviolet Spectroscopic Explorer} \citep[{\it FUSE}, e.g.,][and references therein]{2006ApJ...642..283O}. Variations in the D/H ratio were observed early on \citep{1983ApJ...264..172Y} and by different satellites \citep[e.g.,][]{1999ApJ...520..182J}. \citet{1999NewA....4..231L} discuss some of the mechanisms that could be responsible for the observed variations, based on the D/H data available at the time. However only after the launch of {\it FUSE} in 1999, has a large enough sample of sightlines with differing D/H ratios been accumulated, allowing the question of what causes the D/H variations to be addressed in a more systematic way. While the constancy of the D/H ratio in the Local Bubble \citep[LB, corresponding to log $N$(H) $\leq$ 19.2,][]{1999A&A...346..785S} seems to be a well accepted fact \citep[(D/H)$_{\rm LB}$ = (1.56 $\pm$ 0.04)$\times10^{-5}$ from the compilation by][]{2004ApJ...609..838W} two distinct explanations have emerged to explain the large scatter of gas-phase D/H measurements, in the Solar neighborhood outside of the LB. These measurements range from D/H = (0.50 $\pm$ 0.16)$\times10^{-5}$ toward $\theta$~Car derived by \citet{1992ApJS...83..261A} to D/H = (2.18 $\pm~^{0.22}_{0.19}$)$\times10^{-5}$ toward $\gamma^2$~Vel derived by \citet{2000ApJ...545..277S}.

\citet{2003ApJ...599..297H} proposed a low value for the present-epoch deuterium abundance in the Solar neighborhood, based on indirect measurements of D/H from D/O and the constancy of the O/H ratio \citep[O/H = (3.43 $\pm$ 0.15)$\times10^{-4}$ from][updated $f$-value]{1998ApJ...493..222M}. Using more measurements than \citet{2003ApJ...599..297H}, \citet{2006ASPC..348...47H} proposed then D/H = (0.7 $\pm$ 0.2)$\times10^{-5}$ for the present-epoch deuterium abundance. The high D/H ratios observed would be due to unknown systematic errors in the determination of $N$(H\,I), unknown deuterium enrichment processes and/or local infall of deuterium-rich gas.

The possibility that deuterium could be depleted in dust grains was first proposed by \citet{1982auva.nasa...54J}. \citet{2004AAS...205.5704D} and \citet{2006ASPC..348...58D} explored further the theoretical plausibility of this process leading \citet{2004ApJ...609..838W} and \citet{Linsky2006} to propose that the depletion of deuterium into dust grains leads to different levels of depletion along sightlines with different histories, producing the observed scatter in D/H. In this scenario, the highest measurements of the gas-phase abundance of deuterium in the Solar neighborhood place a lower limit on the total deuterium abundance, D/H $\ge$ (2.31 $\pm$ 0.24)$\times10^{-5}$, which one can then compare to the primordial value through chemical evolution models. Supporting evidence for this model includes correlations of D/H with the depletion of Ti \citep{2005ApJ...620L..39P}, Fe and Si \citep{Linsky2006}, and with the average sightline density \citep{2006ApJ...642..283O}.

Galactic evolution models have to reproduce a number of observables such as the metallicity distribution of late type stars, along with predictions of the evolution of deuterium from its primordial value to its present day distribution in the Galaxy, and in particular its value in the Solar neighborhood. 
 \citet{2006MNRAS.369..295R} have addressed the different views on the behavior of the D/H ratio by considering what value of D/H is predicted for the Solar neighborhood by chemical evolution models that satisfy the majority of the observational constraints (such as the G-dwarf metallicity distribution). They find that a modest star formation and a continuous infall of unprocessed gas is required to fit all the observational data, leading only to a modest decrease of the deuterium abundance from its primordial value. According to the models of \citet{2006MNRAS.369..295R}, which adopt (D/H)$_{\rm P}$ = 2.6$\times10^{-5}$, D/H in the Solar neighborhood should be in the range of (1.4--2.0)$\times10^{-5}$, implying astration factors $<$1.8. Low D/H ratios, as the one proposed by \citet{2006ASPC..348...47H} are ruled out because they produced significant disagreements with other observational constraints, in particular with the G-dwarf metallicity distribution.

Testing the two explanations above for the scatter in D/H, requires more measurements of D/H and other species that might yield information about the depletion and physical conditions along the sightlines, particularly along high column density sightlines. In this work we present analyses of the D/H and other ratios along the extended sightlines to HD\,41161 and HD\,53975. Both of these sightlines have total hydrogen column densities greater than 10$^{21}$ cm$^{-2}$, and according to the variable deuterium depletion model such high column density sightlines should be biased by cold, not recently shocked gas, where deuterium would be depleted in grains. For the HD\,53975 sightline we find D/H = (1.02 $\pm~^{0.23}_{0.20}$)$\times10^{-5}$, a value slightly higher but comparable to those found previously for similar sightlines. The high D/H ratio along the HD\,41161 sightline, D/H = (2.14 $\pm~^{0.51}_{0.43}$)$\times10^{-5}$, presents the first evidence of D/H variations at high $N$(H), indicating that our understanding of the behavior of D/H is still incomplete. 

This paper is organized as follows. The targets and their sightlines are presented in $\S$\ref{targets}. $\S$\ref{obs} describes the observations and data processing, while $\S$\ref{ana} presents the analyses. The abundance of neutral hydrogen along the two sightlines is discussed in $\S$\ref{hi}. The results are presented and discussed in $\S$\ref{results}. Constrains on the $f$-values of Cl\,I transitions present in the spectra of HD\,41161 are derived in $\S$\ref{cl_disc}. Our work is summarized in $\S$\ref{summary}. All uncertainties are quoted at the 1$\sigma$ level unless noted otherwise.

\section{THE TARGETS AND THEIR SIGHTLINES}
\label{targets}

The properties of the two stars are listed in Table \ref{star_properties}. Below we discuss each star in detail.

\subsection{HD\,41161}
\label{hd41161los}

HD\,41161 \citep[O8V,][]{1971ApJ...170..325C} is a field star \citep{2004ApJS..151..103M} that probes gas in the direction $l$~= 164.97$^\circ$~and $b$~= $+$12.89$^\circ$ up to 1253 pc. \citet{1978ApJ...219..845J} used {\it Copernicus} data to derive log $N$(O\,VI) = 13.29. Using {\it IUE} data \citet{1985ApJ...294..599S} derived log $N$(H\,I) = 20.98 $\pm$ 0.07, while \citet{1994ApJS...93..211D} determined log $N$(H\,I) = 21.01 $\pm$ 0.08 also from {\it IUE} data. Taking into account these two determinations of $N$(H\,I) we adopt for the purpose of this study log $N$(H\,I) = 21.00 $\pm$ 0.09.

\citet{1998ApJ...492..569D} determined the column densities of \hmol~($J$ = 0 -- 5) along this sightline using ORFEUS-1 data. The higher resolution and signal-to-noise of the {\it FUSE} data for HD\,41161, allow us to determine \hmol~column densities with uncertainties smaller than the $\sim$60\% (or more) quoted by \citet{1998ApJ...492..569D}.

\subsection{HD\,53975}
\label{hd53975los}

HD\,53975 \citep[O7.5V,][]{1971ApJ...170..325C} is a spectroscopic binary \citep{1980ApJ...242.1063G,1994ApJ...422..823G}, member of the CMa OB1 association \citep{1978ApJS...38..309H} probing gas up to a distance of 1318 pc in the direction $l$~= 225.68$^\circ$~and $b$~= $-$2.32$^\circ$. Based on observations obtained with {\it Copernicus}, \citet{1978ApJ...224..132B} determined log $N$(H\,I) = 21.15 $\pm~^{0.08}_{0.10}$. \citet{1985ApJ...294..599S} used {\it IUE} data to derive log $N$(H\,I) = 21.11 $\pm$ 0.08, while \citet{1994ApJS...93..211D} determined log $N$(H\,I) = 21.10 $\pm$ 0.08, also from {\it IUE} data.  We take the weighted average of the independent $N$(H\,I) determinations based on {\it Copernicus} and {\it IUE} \citep{1985ApJ...294..599S} data to adopt for this work log $N$(H\,I) = 21.13 $\pm$ 0.06.

\section{OBSERVATIONS AND DATA PROCESSING} 
\label{obs}

The {\it FUSE} observatory consists of four coaligned prime-focus telescopes and Rowland-circle spectrographs that produce spectra over the wavelength range 905 -- 1187 \AA, with a spectral resolution of $\sim$15 -- 20 \kms~(wavelength dependent) for point sources. Two of the optical channels employ SiC coatings, providing reflectivity in the wavelength range $\sim$~905~--~1000~\AA,~while the other two have LiF coatings for maximum sensitivity above 1000~\AA. Dispersed light is focused onto two photon-counting microchannel plate detectors. With this arrangement of optical channels (LiF 1, LiF 2, SiC 1, and SiC 2) and detector segments (1A, 1B, 2A, 2B) the \fuse~instrument has 8 channels: LiF 1A, LiF 1B, LiF 2A, LiF 2B, SiC 1A, SiC 1B, SiC 2A, and SiC 2B. Four channels cover the wavelength range 1000~--~1080~\AA~while two channels each cover the ranges 900~--~1000~\AA~ and 1080~--~1180~\AA. Details about the $FUSE$~mission, its planning, and on-orbit performance can be found in \citet{2000ApJ...538L...1M} and \citet{2000ApJ...538L...7S}.

Table \ref{fuse_obs} summarizes the {\it FUSE} observations of HD\,41161 and HD\,53975. HD\,41161 was observed by {\it FUSE} through the large (LWRS, $30\mfarcs\times30\mfarcs$) and medium-sized (MDRS, $4\mfarcs\times20\mfarcs$) apertures. During the short (58 s) LWRS observation data was obtained with the LiF and SiC channels. During the longer (6520 s) MDRS observation only SiC data was obtained due to the flux of this star being very close to the {\it FUSE} bright object limit (10$^{-10}$ erg s$^{-1}$ \AA$^{-1}$ cm$^{-2}$). The lower reflectivity of the SiC channels compared to that of the LiF channels allows some bright targets to be observed, without damaging effects on the detectors. 

The two-dimensional \fuse~spectra are reduced using the CalFUSE pipeline v3\footnote{The CalFUSE pipeline reference guide is available at http://fuse.pha.jhu.edu/analysis/pipeline\_reference.html}. The processing includes data screening for low quality or unreliable data, thermal drift correction, geometric distortion correction, heliocentric velocity correction, dead time correction, wavelength calibration, detection and removal of event bursts, background subtraction, and astigmatism correction. The spectra are aligned by cross-correlating the individual exposures over a short wavelength range that contains prominent spectral features and then coadded by weighting each exposure by its exposure time, using the CORRCAL software developed by S. Friedman.

The S/N of the LWRS data obtained for the HD\,41161 sightline is inferior to that of the MDRS data. In our analysis we use SiC data from the MDRS observation (no LiF data was obtained during this observation) and only LiF data from the LWRS observation. The coadded MDRS data has a signal-to-noise ratio per pixel of 16.87 at $\sim$918.5 \AA~in the SiC 1B channel and of 32.7 at $\sim$1032.2 \AA~in the SiC 1A channel.

HD\,53975 was observed by {\it FUSE} for 482 s through the MDRS aperture. Similarly to HD\,41161 only SiC data were obtained during this observation due to the flux of this star being close to the {\it FUSE} bright object limit. We coadded the data for this target in the manner described above. The coadded data has a S/N per pixel of 10.25 at $\sim$ 918.5 \AA~in the SiC 1B channel and of 16.87 at $\sim$1032.6 \AA~in the SiC 1A channel.

\begin{figure}
\begin{center}
\epsscale{0.87}
\plotone{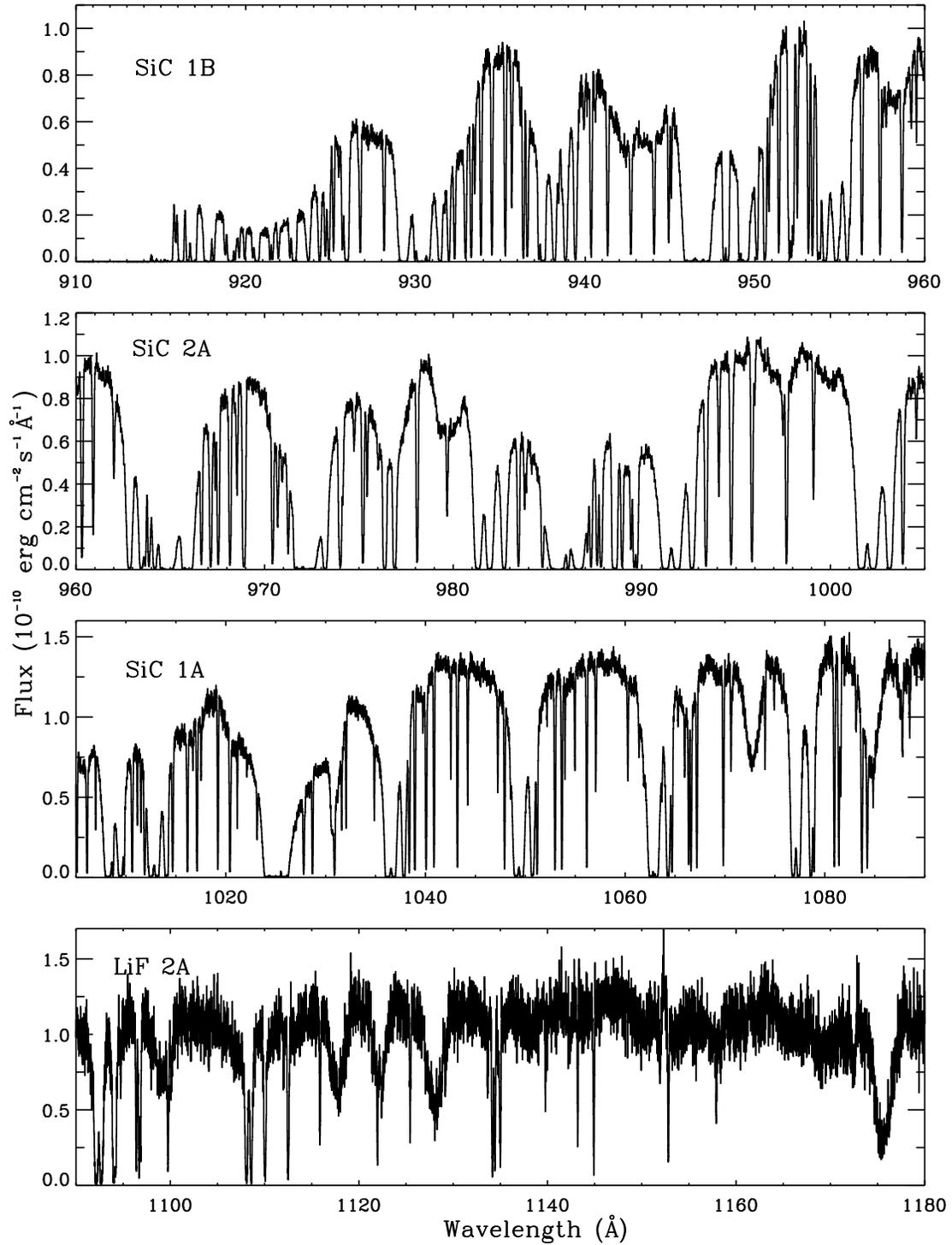}
\caption{$FUSE$~spectrum of HD\,41161. The {\it FUSE} channel used for each panel is indicated at the top, in the left. The SiC data displayed is from the MDRS observation, the LiF one from the LWRS observation (no LiF data obtained in the MDRS observation; see Table \ref{fuse_obs}). \label{hd41161spectra}}
\end{center}
\end{figure}

\begin{figure}
\begin{center}
\epsscale{0.87}
\plotone{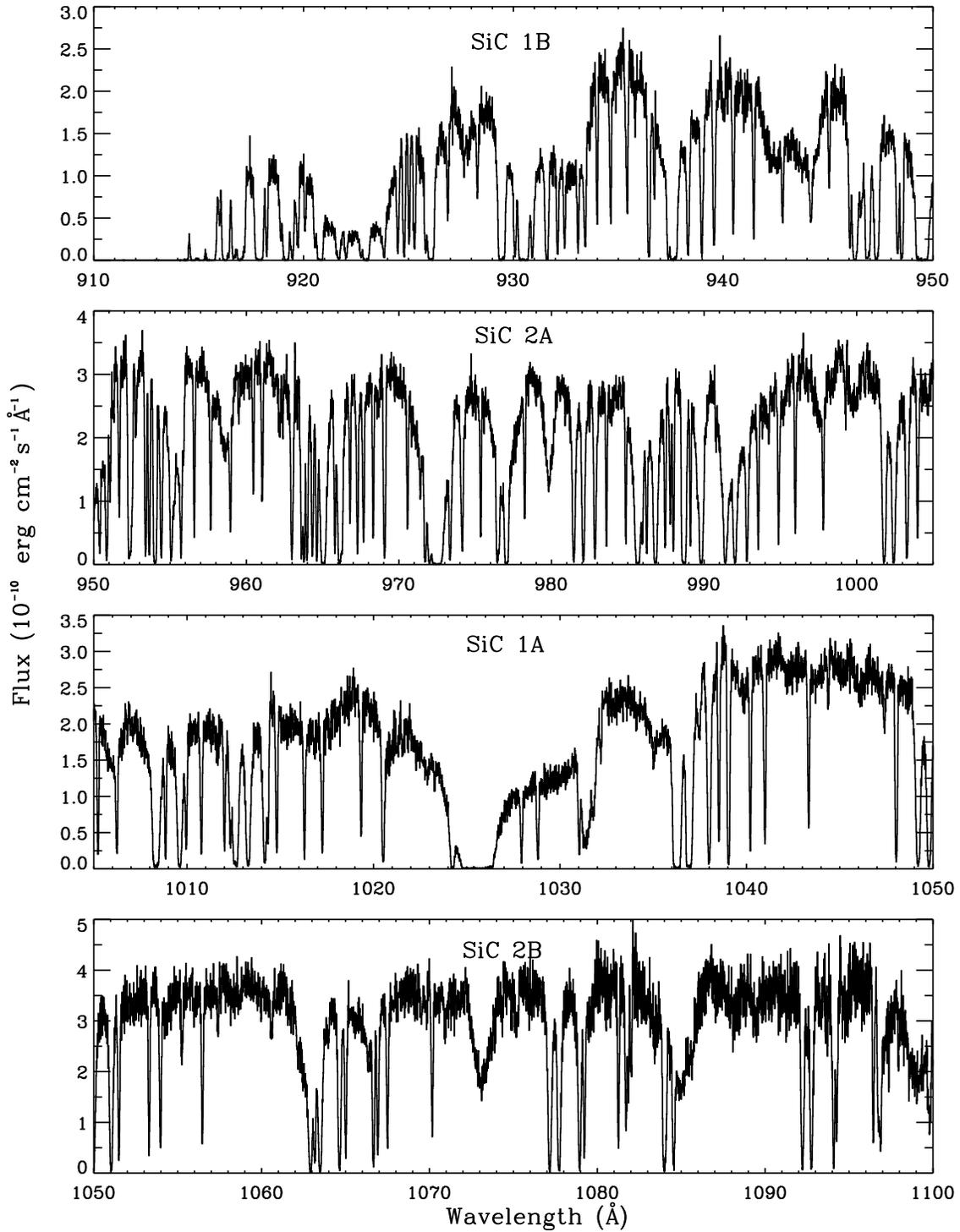}
\caption{$FUSE$~spectrum of HD\,53975. The {\it FUSE} channel used for each panel is indicated at the top, in the center. No LiF data was obtained for this target. \label{hd53975spectra}}
\end{center}
\end{figure}

Figure \ref{hd41161spectra} and \ref{hd53975spectra} present the {\it FUSE} spectra of the two targets.

\section{ANALYSIS}
\label{ana}

We use apparent optical depth, curve of growth, and profile fitting methods (AOD, COG, and PF, respectively) to determine column densities along the sight lines, whenever possible \citep[see e.g.][for further discussion of these methods]{2003ApJ...587..235O,2006ApJ...642..283O}. Details about the particular fitting routine used here ({\it Owens.f}) can be found in \citet{2002ApJS..140..103H} and \citet{2002ApJS..140...67L}.

Table \ref{atomicdata} lists the atomic data and equivalent widths of the transitions used in this study. A, C, and P are used to indicate which transitions are used with the AOD, COG, and PF techniques, respectively. No equivalent widths were measured for the transitions flagged only with 'P' due to blendings or in the case of D\,I because there are not enough unblended D\,I transitions to construct a curve-of-growth. These transitions were only used with the PF technique. We use the compilation by \citet{2003ApJS..149..205M} for the atomic data and that of \citet{1993A&AS..101..273A} and \citet{1993A&AS..101..323A} for the molecular data.

\subsection{Atomic species along the line of sight to HD\,41161}
\label{hd41161ana}

We use the AOD technique to place lower limits on the column densities of C\,I, C\,III, N\,II, P\,II, Cl\,II, and Ar\,I along this sight line and to determine the column densities of D\,I, N\,I, Fe\,II. With the COG method we derive column densities for N\,I, Fe\,II and $b_{\rm N\,I}$ = 8.3 $\pm$ 0.2 km s$^{-1}$ and $b_{\rm Fe\,II}$ = 9.2 $\pm~^{1.0}_{0.9}$ km s$^{-1}$. 

\begin{figure}
\begin{center}
\epsscale{0.75}
\plotone{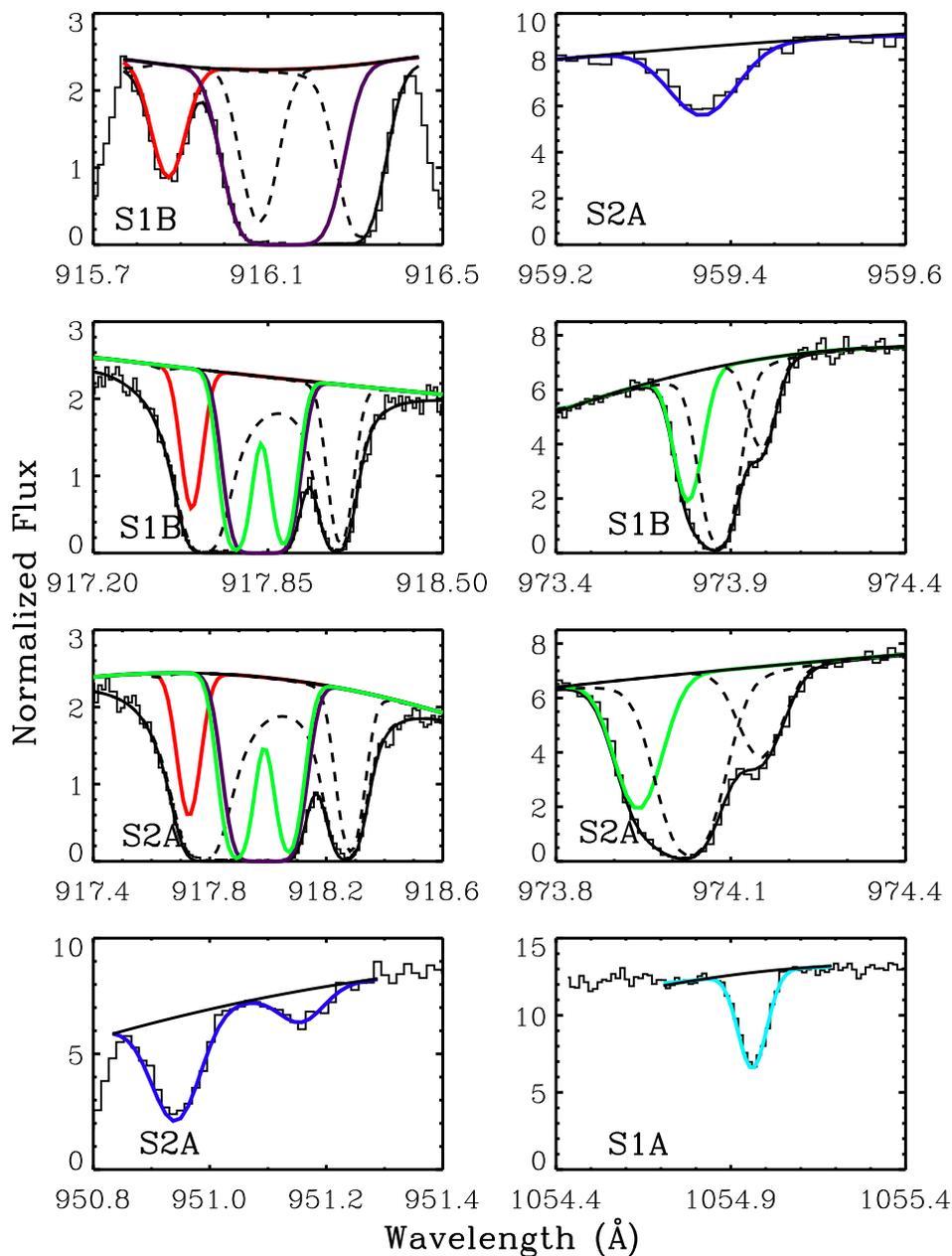}
\caption{Fits to some of the lines used to determine column densities along the HD\,41161 sightline (convolved with the instrument LSF). Absorption by H\,I is representd by magenta, D\,I by red, N\,I by dark blue, O\,I by green, and Fe\,II by cyan. Absorption by \hmol~is represented by black dashed lines. See $\S$ \ref{hd41161ana} for discussion. \label{hd41161fits}}
\end{center}
\end{figure}

The PF technique is used in a similar manner in the analyses of both sightlines. We fit a single absorption component to the low ionization atomic species D\,I, N\,I, O\,I, and Fe\,II, and another to the molecular species. H\,I is included in the fit as well, in a separate absorption component, with the sole purpose of modeling the continuum in the vicinity of the D\,I lines. We use a line spread function with a FWHM of 0.0641 \AA~to fit the {\it FUSE} data, constant for all channels and wavelengths, corresponding to a spectral resolution of $\sim$20 km s$^{-1}$ \citep[see][for a determination of the LSF across the {\it FUSE} bandpass and for a comparison between $N$ obtained both with free and fixed LSFs]{2005ApJ...625..210W}. We derive a Doppler parameter $b$ $\sim$ 10 km s$^{-1}$ for the absorption component containing D\,I, N\,I, O\,I, and Fe\,II, in agreement with $b$ derived for N\,I and Fe\,II with the COG technique. This Doppler parameter should not be viewed as the classic $b$ composed of turbulent and thermal terms, but rather the result of line broadening due to multiple unresolved components along the sightline. Fits to some of the lines used with the PF method are presented in Figure \ref{hd41161fits}. $N$(D\,I) along this sightline is constrained mostly by the $\lambda$916.2 D\,I transition, which is free from blends with other lines, making $N$(D\,I) independent of $N$(\hmol).

Table \ref{NHD41161} presents the column densities of the atomic species obtained with the different methods as well as the adopted values. In this work we adopt values that are roughly at the midpoint of the values obtained with the different methods and uncertainties that include the extremes of the values obtained with the different methods. As a general rule we do not adopt 1$\sigma$ uncertainties smaller that 0.05 dex ($\sim$12\%) because we feel that adopting smaller uncertainties could be misleading, due to the presence of unknown systematic effects.

O\,VI is not clearly detected in our data. Using the $\lambda$1032 O\,VI transition we place a 3$\sigma$ upper limit on log $N$(O\,VI) of 13.0, while \citet{1978ApJ...219..845J} determined log $N$(O\,VI) = 13.29 using {\it Copernicus} data. Since no uncertainties were quoted by \citet{1978ApJ...219..845J} and given the S/N of the {\it Copernicus} data near the O\,VI $\lambda$1032 transition it is likely that the two determinations are compatible.

Cl\,I is also detected along the HD\,41161 sightline. We use the $\lambda$1088 Cl\,I transition to determine $N$(Cl\,I) along this sight line with the AOD method. Besides the $\lambda$1088 transition, many other Cl\,I transitions (for which only theoretically derived oscillator strengths are available) are detected. The high S/N ratio of the data allows us to place constraints on the $f$-values of several of these transitions. This is discussed in detail in $\S$ \ref{cl_disc}.

\subsection{\hmol~and HD along the line of sight to HD\,41161}
\label{h2disc_hd41161}

Absorption by molecular hydrogen from the rotational levels $J$ = 0 through $J$ = 5 and by HD ($J$ = 0) is present in the spectra of HD\,41161. We use the AOD technique to determine the column densities of HD ($J$ = 0) and \hmol~($J$ = 4 and $J$ = 5).

We fit a single-$b$ COG to the measured equivalent widths of HD and \hmol~($J$ = 1 through 5) for the HD\,41161 sight line. Transitions from the rotational level $J$ = 0 are too blended with those of other species to allow measurements of equivalent widths. We determine $b_{\rm H_2}$ = 7.2 $\pm~^{0.4}_{0.2}$ km s$^{-1}$.

We use the PF technique to determine the column densities of HD and \hmol~along this sightline, by fitting one absorption component simultaneously, to multiple transitions of HD and \hmol. We derive $b_{\rm H_2}$ $\sim$ 7 km s$^{-1}$.
  
Table \ref{h2col} presents the adopted molecular column densities along this sightline (following the same reasoning above, used to adopt the atomic column densities for this sightline). Our results agree with those derived by \citet{1998ApJ...492..569D} except for $J$ = 2, for which they quote log$N$($J$ = 2) = 17.6 $\pm$ 0.2, while we find log$N$($J$ = 2) = 17.97 $\pm$ 0.10. We find log $N$(\hmol)~=~19.98 $\pm$ 0.08. Using the column densities for the $J$ = 0 and $J$ = 1 levels we derive the excitation temperature $T_{\rm 01}$ = 84 $\pm$ 7 K. This temperature is in good agreement with that found by \citet{1977ApJ...216..291S} from {\it Copernicus} observations of 61 sightlines with log $N$(\hmol) $>$ 18.0 (77 $\pm$ 17 K) and that derived by \citet{2002ApJ...577..221R} from a {\it FUSE} survey of sightlines with $A_v \geq$ 1 mag (68 $\pm$ 15 K).

We derive log $N$(\hmol) = 19.98 $\pm~^{0.08}_{0.09}$, which combined with $N$(H\,I) from $\S$\ref{hd41161los} leads to the total hydrogen column density along this sightline log $N$(H) = log ($N$(H\,I) + 2$N$(\hmol)) = 21.08 $\pm$ 0.08, and to the molecular fraction $f_{\rm H_2}$ = 2$N$(\hmol)/$N$(H) = (16.0~$\pm~^{4.0}_{3.4}$) \%.

\subsection{Atomic species along the line of sight to HD\,53975}
\label{hd53975ana}

The AOD technique is used to place lower limits on the column densities of C\,I, C\,III, N\,II, P\,II, S\,III, and Ar\,I and to determine column densities of D\,I, N\,I, O\,VI, Fe\,II, and Cl\,I along the HD\,53975 sight line. 

We use the COG method to derive column densities of N\,I and Fe\,II along the HD\,53975 sightline. We derive $b_{\rm N\,I}$ = 9.6 $\pm$ 0.4 km s$^{-1}$ and $b_{\rm Fe\,II}$ $\sim$ 12 km s$^{-1}$.

\begin{figure}
\begin{center}
\epsscale{0.75}
\plotone{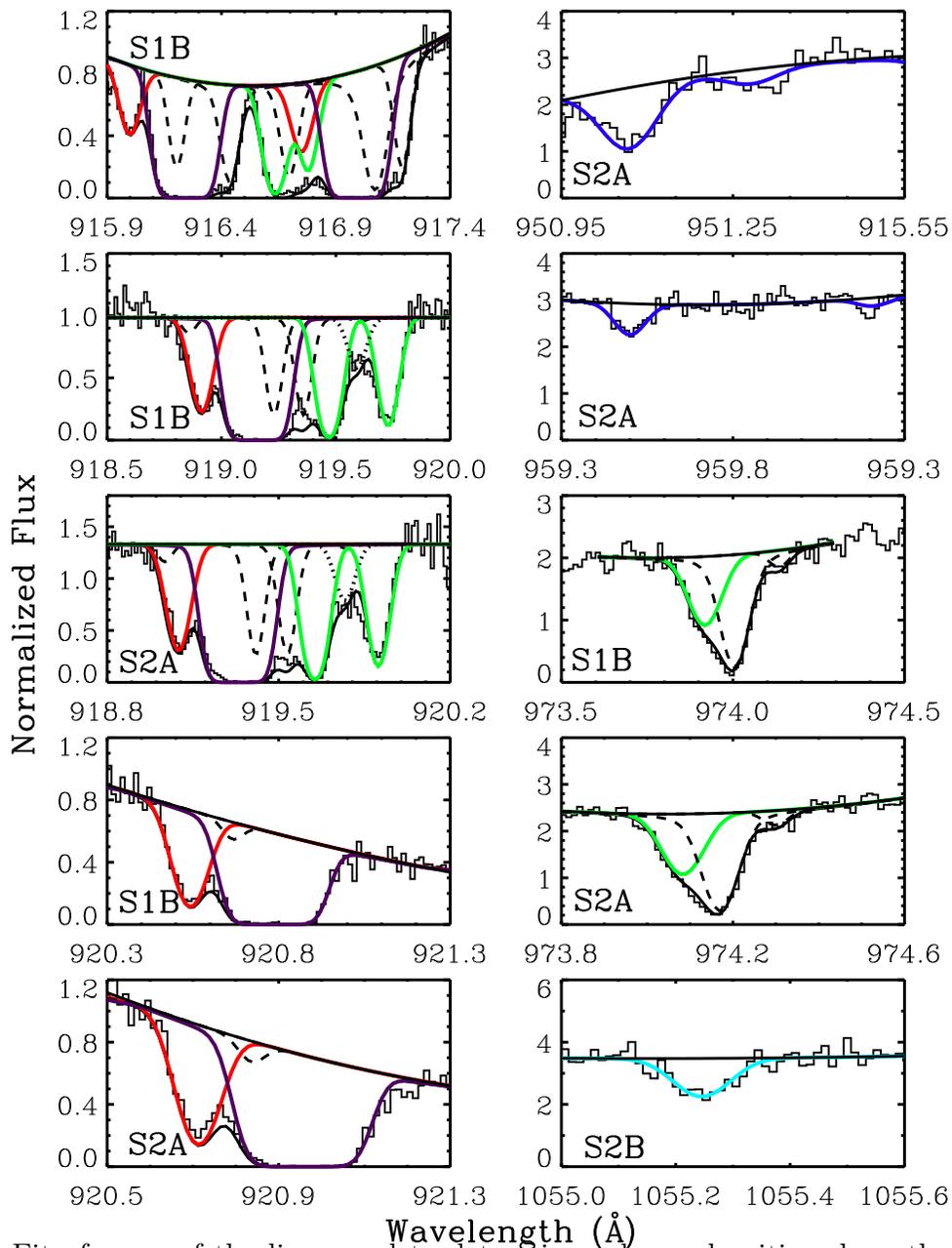}
\caption{Fit of some of the lines used to determine column densities along the HD\,53975 sightline (convolved with the instrument LSF). Colors are the same as in Figure \ref{hd41161fits}. Absorption by Ar\,II is represented by a black dotted line. See $\S$ \ref{hd53975ana} for discussion. \label{hd53975fits}}
\end{center}
\end{figure}

With the PF technique we fit a single absorption component to the low ionization atomic species D\,I, N\,I, O\,I, Fe\,II, and Mg\,II, and another to the molecular species. H\,I is included in the fit as well, in a separate absorption component, with the sole purpose of modeling the continuum in the vicinity of the D\,I lines. A separate component is used to model the absorption by O\,VI.
We use a line spread function with a FWHM of 0.0641 \AA~to fit the {\it FUSE} data, constant for all channels and wavelengths, corresponding to a spectral resolution of $\sim$20 km s$^{-1}$. Figure \ref{hd53975fits} presents fits to some of the lines used to determine the column densities along this sightline. We derive $b$ $\sim$ 15 km s$^{-1}$ for the absorption component containing D\,I. Similarly to HD\,41161, this Doppler parameter reflects the broadening of the lines due to multiple unresolved absorption components along the line of sight. From the 4 D\,I transitions used to constrain $N$(D\,I) (see Table \ref{atomicdata}) only the $\lambda$ 916.9 line suffers from major blends with \hmol~($J$ = 3) whose column density is not well constrained (see $\S$ \ref{h2disc_hd53975} below). Removing this D\,I line from the fits does not affect $N$(D\,I) adopted. Table \ref{NHD53975} presents the atomic column densities along this sightline obtained with the different methods as well as the adopted values (see $\S$ \ref{hd41161ana} for how values and uncertainties are adopted).

\subsection{\hmol~and HD along the line of sight to HD\,53975}
\label{h2disc_hd53975}

Absorption by HD and by molecular hydrogen from the rotational levels $J$ = 0 through $J$ = 5 is detected along this sightline. We use the AOD technique to determine the column densities of the \hmol~$J$ = 4 and $J$ = 5 levels.

Similarly to HD\,41161 we fit a single-$b$ COG to the measured equivalent widths of \hmol ($J$ = 0 through 5) and we use the PF method to determine the column densities of the different \hmol~$J$~levels and HD by fitting one absorption component to \hmol and HD. We find that the column densities of the levels $J$ = 0 through $J$ =3 derived with the COG method are at least a factor of $\sim$ 2 larger than the ones derived with the PF technique. An \hmol~ absorption model containing the column densities and Doppler parameter derived with the single-$b$ COG leads to \hmol~absorption profiles that are inconsistent with the data and that clearly overestimate the true $N$(\hmol) along this sightline. For this reason we adopt for $N$(\hmol) along this sightline the results derived with the AOD and PF techniques. Transitions from the $J$ = 2 and $J$ = 3 rotational levels are on the flat part of the curve of growth, hence the associated column densities are very uncertain and are flagged with ':' in Table \ref{h2col}.

We note that $N$(D\,I) is not affected by the adopted $N$(\hmol) as fits performed where the D\,I lines blended with \hmol~were removed from the fit (particularly $\lambda$916.9) lead to $N$(D\,I) consistent with the value adopted in Table \ref{NHD53975}.

Table \ref{h2col} presents the adopted molecular column densities for this sightline (see $\S$ \ref{hd41161ana} for how values and uncertainties are adopted). We find log $N$(\hmol)~=~19.18 $\pm$ 0.04 by adding the column densities of the $J$ = 0 and $J$ = 1 levels, which contain most of the \hmol~along this sightline. Using the column densities for these two levels we derive the excitation temperature $T_{\rm 01}$ = 93 $\pm$ 9 K, in agreement with that found by \citet{1977ApJ...216..291S} but slightly higher than that found by \citet{2002ApJ...577..221R}. The total hydrogen column density along this sightline is then $N$(H) = 21.14 $\pm$ 0.06 with $f_{\rm H_2}$ = (2.2 $\pm~^{0.4}_{0.3}$)\%.

\section{NEUTRAL HYDROGEN ABUNDANCES}
\label{hi}

\subsection{$N$(H\,I) ALONG THE HD\,41161 LINE OF SIGHT}
\label{hihd41161}

\begin{figure}
\begin{center}
\epsscale{0.6}
\plotone{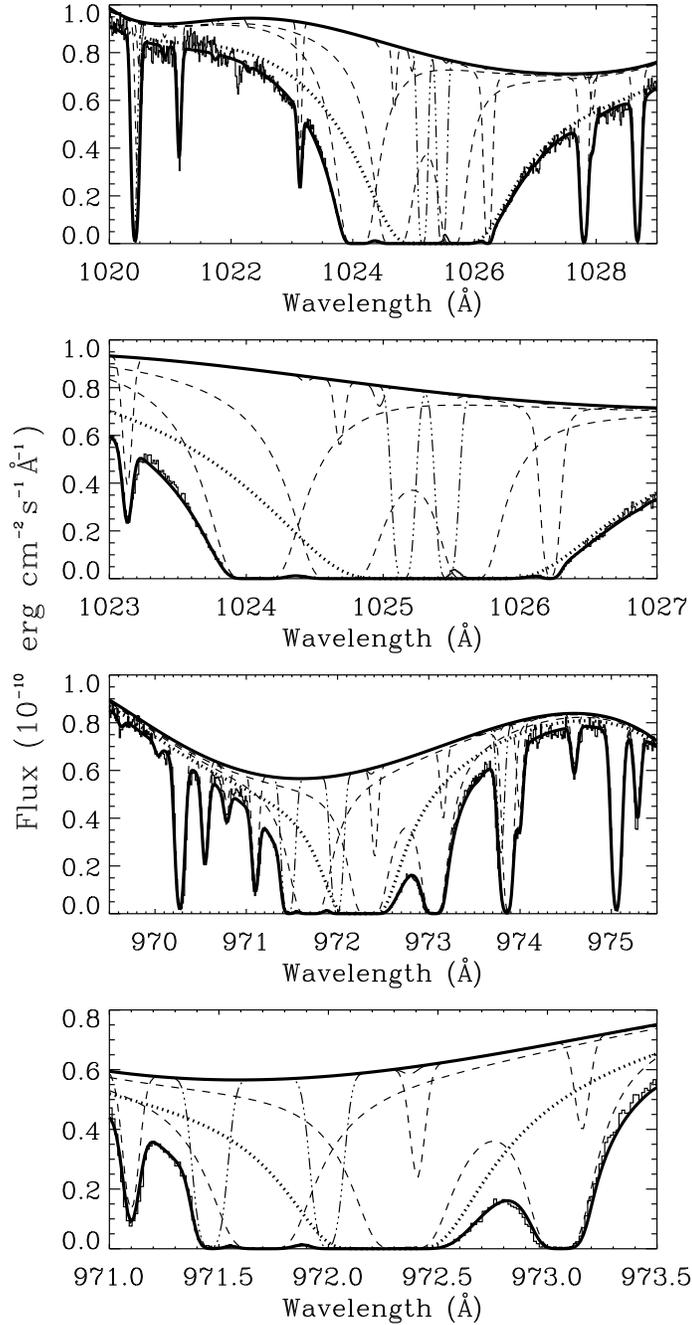}
\caption{Fit to H\,I Ly$\beta$ (first and second panels) and Ly$\gamma$ (third and fourth panels) transitions along the HD\,41161 sightline, yielding log $N$(H\,I) = 21.13 $\pm~^{0.04}_{0.05}$, in agreement with the value adopted in \S\ref{hd41161los}. The cores of the Ly$\beta$ and Ly$\gamma$ lines are presented in more detail in the second and fourth panels, respectively. \hmol~is represented by dashed lines, O\,I, Si\,II, D\,I, and Fe\,II by dash-dot-dot. H\,I is represented by a thick dotted line. The two top panels are from S1A MDRS, while the bottom two are from S2B MDRS. The small bump near 1025.5 \AA~is due to airglow emission, and does not affect our measurements. See \S\ref{hihd41161} for discussion. \label{hifit_hd41161}}
\end{center}
\end{figure}

To check the consistency of $N$(H\,I) determined from the Ly$\alpha$ transition in the {\it IUE} data, we use some of the H\,I Lyman transitions in the {\it FUSE} bandpass to determine $N$(H\,I), with the profile fitting technique. The following transitions are used to constrain $N$(H\,I): $\lambda\lambda$ 1025, 972, 920, 918, 917, and 916. No stellar model is used in conjunction with the {\it FUSE} data. We fit an absorption model with one component of \hmol, one component containing D\,I, O\,I, Fe\,II, and Ar\,I, and another component with H\,I. The damped wings of the Ly$\beta$ transition constrain $N$(H\,I), while the weaker Lyman lines constrain $b_{\rm H\,I}$. In addition to fitting the Lyman lines we include also other regions of the spectra containing absorption lines of \hmol~and other atomic species, to constrain the shapes of absorption lines of these species that are unresolved from the H\,I Lyman transitions. Figure \ref{hifit_hd41161} presents the fit to the Ly$\beta$ and Ly$\gamma$ H\,I transitions along the HD\,41161 sight line. The fit yields log $N$(H\,I) = 21.13 $\pm~^{0.04}_{0.05}$ and $b_{\rm H\,I}$ $\sim$ 14 km s$^{-1}$, in reasonable agreement with log $N$(H\,I) = 21.00 $\pm$ 0.09 adopted (see \S\ref{hd41161los}). Varying the degree of the polynomials used to define the continua for the Ly$\beta$ and Ly$\gamma$ transitions leads to log $N$(H\,I) in agreement with 21.13 $\pm~^{0.04}_{0.05}$.

\subsection{$N$(H\,I) ALONG THE HD\,53975 LINE OF SIGHT}
\label{hihd53975}

\begin{figure}
\begin{center}
\epsscale{0.6}
\plotone{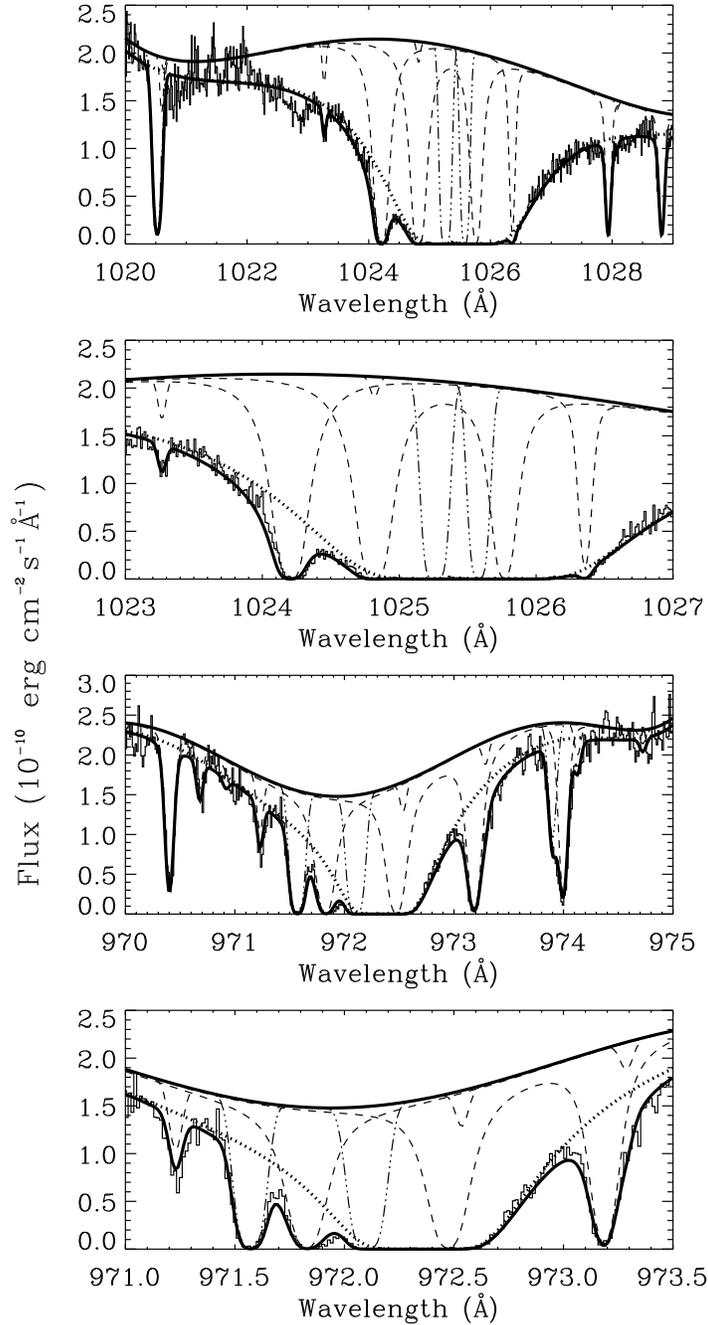}
\caption{Same as Fig. \ref{hifit_hd41161} but for the HD\,53975 sightline, yielding log $N$(H\,I) = 21.19 $\pm$ 0.01, in agreement with the value adopted in \S\ref{hd53975los}. The two top panels are from S1B, while the bottom two are from S1A. See \S\ref{hihd53975} for discussion.\label{hi21.19_hd53975}}
\end{center}
\end{figure}

Similarly to HD\,41161, we check the consistency of $N$(H\,I) derived from the {\it IUE} and {\it Copernicus} data by fitting some of the Ly H\,I transitions in the {\it FUSE} bandpass. We fit an absorption model with one component of \hmol, a second component with D\,I, N\,I, O\,I, and Fe\,II, and a third component containing H\,I. The following H\,I Lyman lines are used to constrain $N$(H\,I): $\lambda\lambda$ 916.4, 917.2, 919.3, 920.9, 972, and 1025. Figure \ref{hi21.19_hd53975} presents the resulting fits to the H\,I Ly$\beta$ and Ly$\gamma$ transitions. We derive log $N$(H\,I) = 21.19 $\pm$ 0.01 (formal statistical errors) and $b_{\rm H\,I}$ $\sim$ 15 km s$^{-1}$, in agreement with the adopted value log $N$(H\,I) = 21.13 $\pm$ 0.06. 




\section{RESULTS AND DISCUSSION}
\label{results}

Table \ref{ratios} summarizes several ratios along the two sightlines. Below we discuss each ratio in detail. 

\subsection{D/H ratios}

In the discussion below the D/H ratios are determined using D/H = [$N$(D) $+$ $N$(HD)]/[$N$(H\,I) $+$ 2$\times$$N$(\hmol) $+$ $N$(HD)].

\subsubsection{D/H along the HD\,41161 sightline}
\label{highhihd41161}

\begin{figure}
\begin{center}
\epsscale{0.75}
\rotatebox{90}{
\plotone{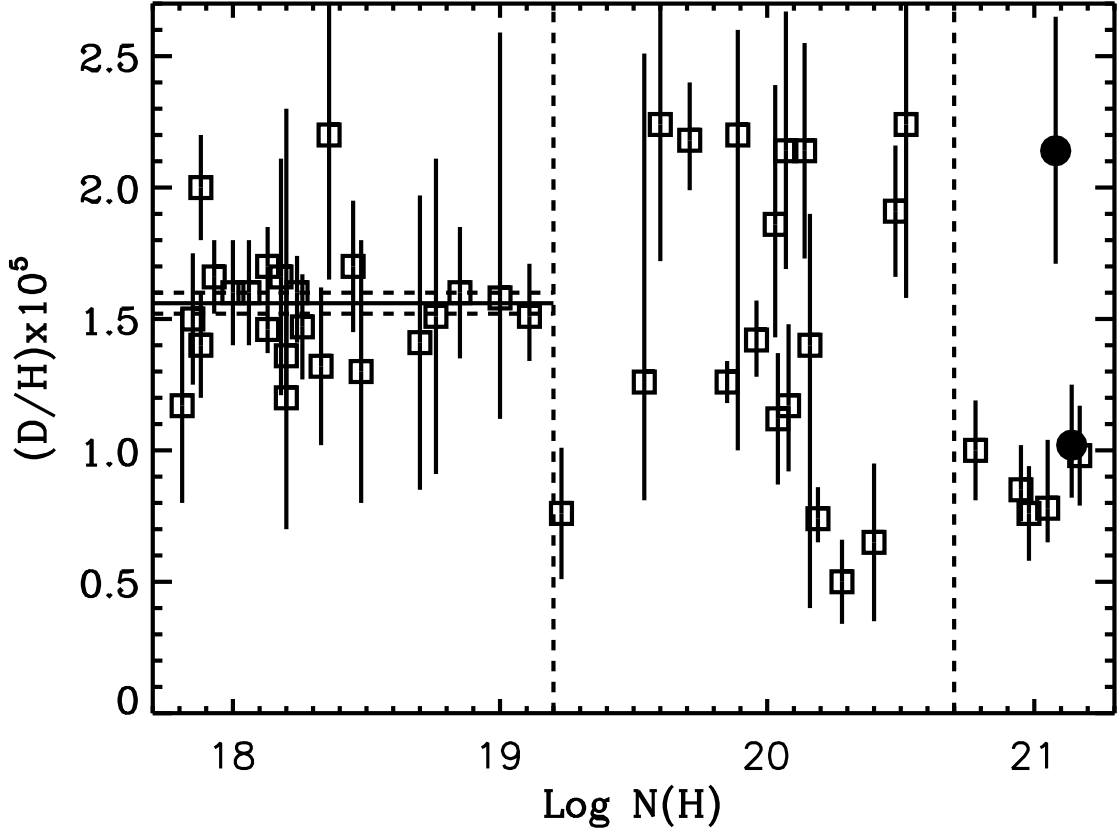}
}
\caption{D/H as a function of log $N$(H). Open squares represent ratios from the literature and from \citet{2006ApJ...642..283O} (where references to all the individual ratios presented can be found), filled circles the two new ratios derived in this work. Dashed and solid horizontal lines mark the Local Bubble D/H ratio (1.56 $\pm$ 0.04)$\times10^{-5}$ derived from a compilation of measurements by \citet{2004ApJ...609..838W}. Dashed vertical lines mark log $N$(H) = 19.2 corresponding to the edge of the Local Bubble, and log $N$(H) = 20.7 after which \citet{2004ApJ...609..838W} and \citet{Linsky2006} proposed that D/H was low and constant. \label{dhratioplot}}
\end{center}
\end{figure}

Figure \ref{dhratioplot} presents the D/H ratio as a function of log $N$(H). Open squares represent values from the literature and the three new values from \citet{2006ApJ...642..283O} (references for all the ratios can be found in this paper), filled circles the ratios derived in this work. The Local Bubble D/H ratio and uncertainties are represented by horizontal solid and dashed lines. The vertical dashed line at log $N$(H) = 19.2 represents the edge of the Local Bubble, while the vertical dashed line at log $N$(H) = 20.7 represents the approximate position of where a new low and constant D/H regime was proposed to exist by \citet{2004ApJ...609..838W} and \citet{Linsky2006}.

\begin{figure}
\begin{center}
\epsscale{0.6}
\plotone{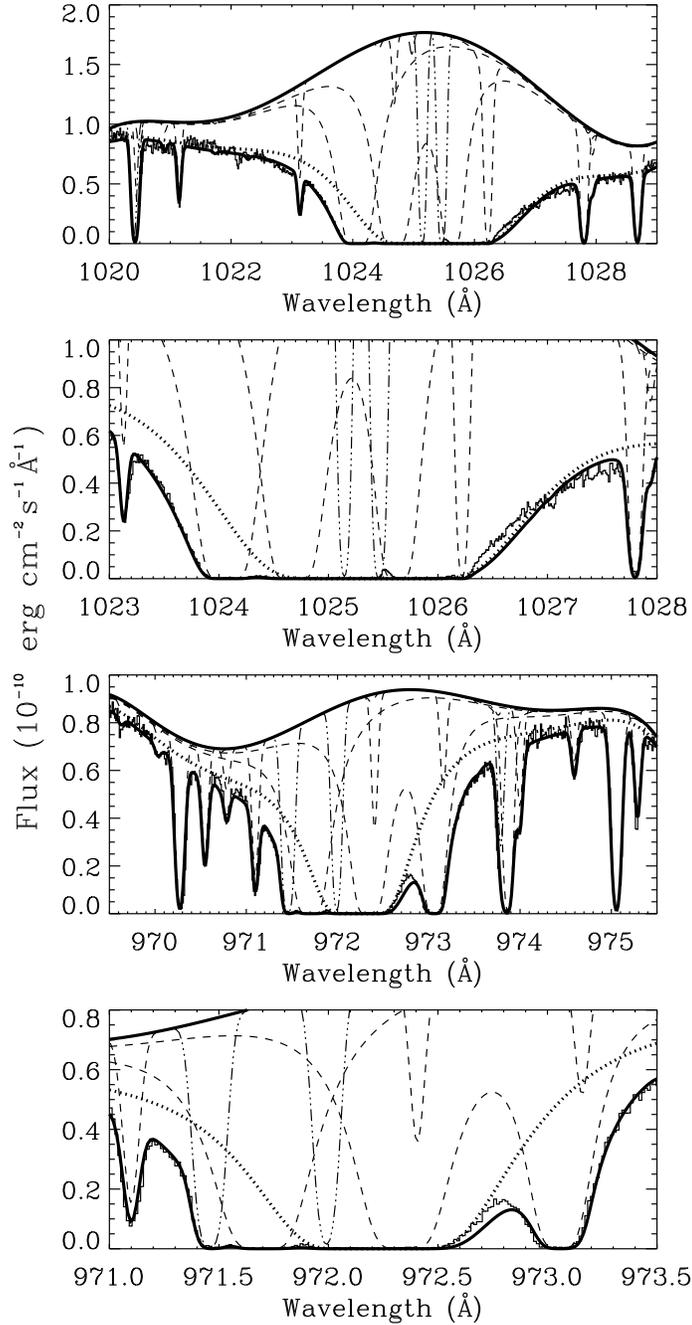}
\caption{Forced fit to H\,I Ly$\beta$ (first and second panels) and Ly$\gamma$ (third and fourth panels) transitions along the HD\,41161 sightline, with log $N$(H\,I) = 21.44, required to bring D/H along this sightline to agree with D/H for other 5 sightlines with similar N(H). Dashed lines represent \hmol, dash-dot-dot O\,I, Si\,II, D\,I, and Fe\,II. H\,I is represented by a thick dotted line. The two top panels are from S1A MDRS, while the bottom two are from S2B MDRS. The cores of the Ly$\beta$ and Ly$\gamma$ transitions, presented in more detail in the second and fourth panels, show that log $N$(H\,I) = 21.44 clearly overestimates the true H\,I absorption. See \S\ref{highhihd41161} for discussion. \label{hi21.43_hd41161}}
\end{center}
\end{figure}

\begin{figure}
\begin{center}
\epsscale{0.7}
\rotatebox{90}{
\plotone{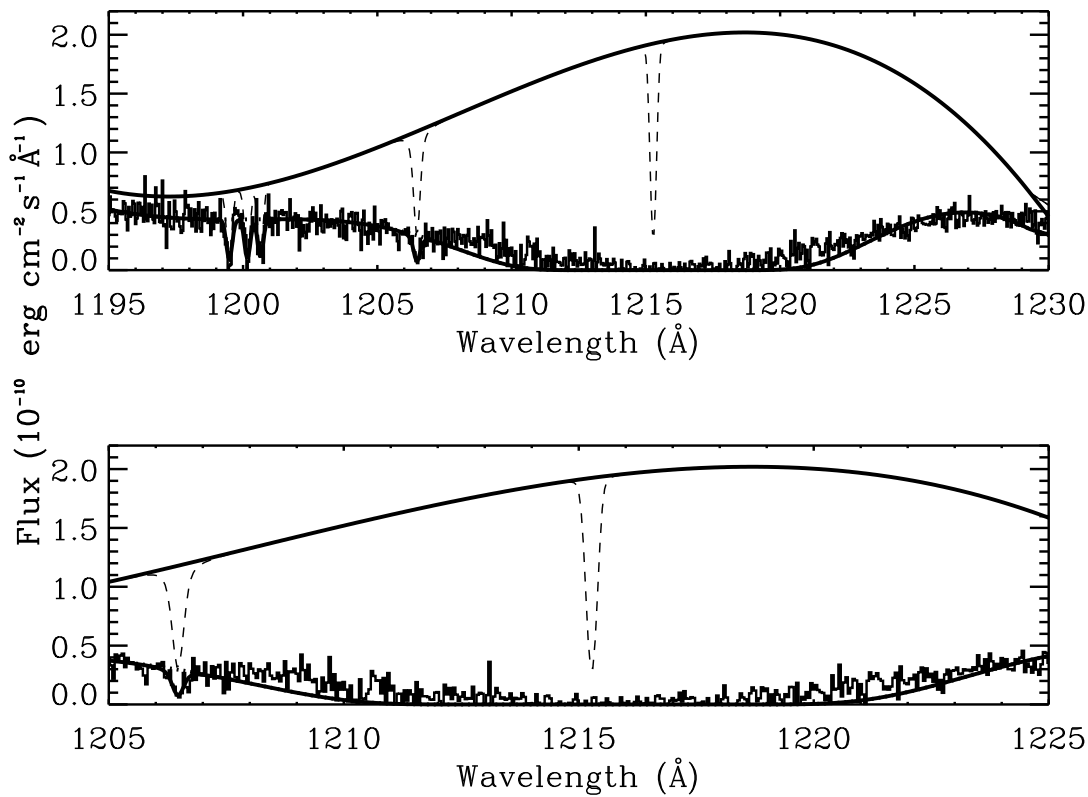}
}
\caption{Forced fit to the H\,I Ly$\alpha$ transition along the HD\,41161 sightline, with log $N$(H\,I) = 21.44, required to bring D/H along this sightline to agree with D/H for other 5 sightlines with similar N(H). Dashed lines represent N\,I, Si\,III, and D\,I. The global fit, including H\,I, as well as the continuum, are represented by thick solid lines. The core of the Ly$\alpha$ transition, presented in more detail in the second panel clearly shows that log $N$(H\,I) = 21.44 clearly overestimates the true H\,I absorption. See \S\ref{highhihd41161} for discussion. \label{hi21.44iue}}
\end{center}
\end{figure}

There are five sight lines with log $N$(H) $>$ 20.7, with D/H ratios available in the literature \citep[HD\,198965, HD\,191877, JL\,9, LS\,1274, and HD\,90087;][]{2003ApJ...586.1094H,2004ApJ...609..838W,2005ApJ...635.1136H}. For these sightlines with 20.7 $<$ log $N$(H) $<$ 21.20, the weighted mean is D/H = (0.86 $\pm$ 0.08)$\times10^{-5}$. HD\,41161 has $N$(H) similar to these sightlines, but a D/H ratio inconsistent by $\sim$ 3$\sigma$. To make D/H along the HD\,41161 sightline consistent with the weighted ratio of the other 5 sightlines, one would have to increase $N$(H) by a factor of $\sim$ 2.49, or log $N$(H\,I) from 21.00 to $\sim$ 21.44. To test whether D/H along this sightline could be high due to $N$(H\,I) being underestimated we performed fits where log $N$(H\,I) was forced to 21.44. Figure \ref{hi21.43_hd41161} presents the fits to H\,I Ly$\beta$ and Ly$\gamma$ with log $N$(H\,I) forced to 21.44. The degree of the polynomial used to fit the continuum of these two transitions, is the same as the one used for the fit where we determined $N$(H\,I) = 21.13 $\pm~^{0.04}_{0.05}$. However, in this case, the continuum is forced to be artificially high to accommodate the higher $N$(H\,I). In addition, the fit grossly overestimates the absorption on the blue side of the core of the Ly lines as shown in the two bottom panels of Figure \ref{hi21.43_hd41161}. To further check that log $N$(H\,I) = 21.44 overestimates the true H\,I absorption we retrieved the {\it IUE} Ly$\alpha$ spectrum for the HD\,41161 sightline from the MAST archive (no further processing was applied). Figure \ref{hi21.44iue} presents the fit to H\,I Ly$\alpha$ forcing log $N$(H\,I) = 21.44. Similarly to the fits to Ly$\beta$ and Ly$\gamma$ discussed above, forcing such a high value of $N$(H\,I) on the data forces the continuum to be artificially high and overestimates the absorption near the core of the Ly$\alpha$ line.

The $N$(H\,I) adopted for this sightline is based on measurements performed by two different sets of authors \citep{1982ApJS...48...95S,1994ApJ...427..274D} who obtain similar results, although both used the same {\it IUE} dataset. In addition, fits of the Ly$\beta$ and Ly$\gamma$ H\,I transitions produce $N$(H\,I) in agreement with the value adopted in $\S$\ref{hd41161los}, while forcing $N$(H\,I) to the value required to bring D/H to agree with the other low D/H measurements in the literature (discussed above) produces fits that clearly overestimate the H\,I absorption. We conclude then that the high D/H along this sightline is unlikely due to $N$(H\,I) being underestimated. 

\begin{figure}
\begin{center}
\epsscale{0.6}
\rotatebox{90}{
\plotone{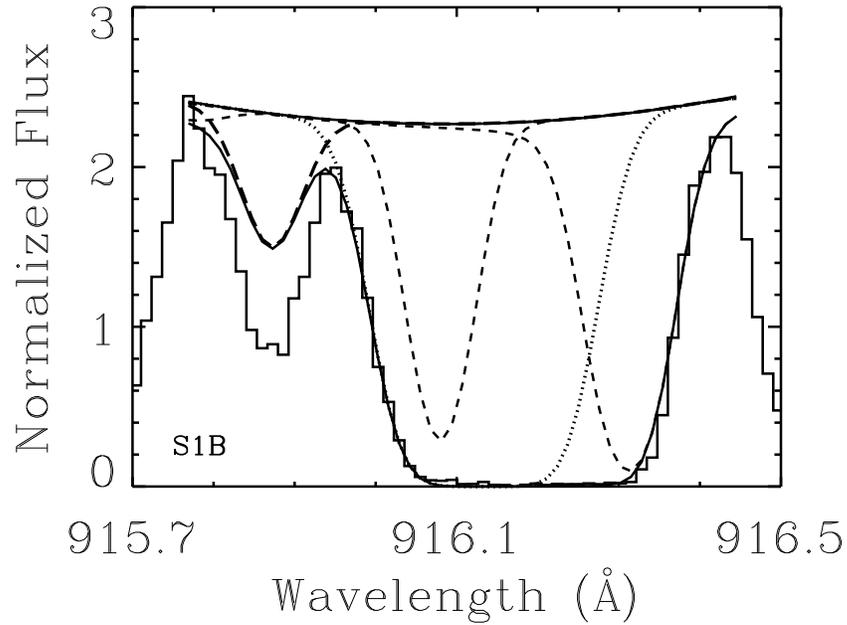}
}
\caption{Log $N$(D\,I) = 16.00, required to lower D/H along the HD\,41161 sightline, superimposed on D\,I $\lambda$916.2. Absorption is represented by short-dashed lines for H\,I, long-dashed lines for D\,I, and dotted lines for \hmol. The overall fit and continuum are represented by solid lines. This figure shows that log $N$(D\,I) = 16.00 clearly underestimates the true D\,I absorption. Compare this figure to the corresponding panel in Fig. \ref{hd41161fits} where we derive log $N$(D\,I) = 16.43 $\pm$ 0.03. See \S\ref{highhihd41161} for discussion. \label{di16}}
\end{center}
\end{figure}

Another possibility to explain the high D/H along this sightline is that $N$(D\,I) is overestimated. This is probably unlikely since saturation tends to move column densities in the opposite direction, i.e., underestimation rather than overestimation. However other factors, such as continuum placement, could produce the desired effect. A D\,I column density of log $N$(D\,I) = 16.00 is required to bring D/H along this sightline down to the weighted mean derived by the 5 sightlines discussed above. To test whether this $N$(D\,I) value is consistent with the data we consider the $\lambda$916.18 and $\lambda$919.1 D\,I transitions. With the AOD method we derive log $N$(D\,I) = 16.37 $\pm$ 0.04 from the $\lambda$916.2 transition and our fits yield log $N$(D\,I) = 16.43 $\pm$ 0.03. A column density of log $N$(D\,I) = 16.0 clearly underestimates the true D\,I absorption, as shown in Figure \ref{di16} where we overplot the above column density on the $\lambda$916.2 transition. 

To further check our $N$(D\,I) measurement along this sightline we can consider the $\lambda$919.1 D\,I transition. This transition, not used in our analysis of this sightline, is blended with \hmol~($J$ = 4) and according to our model slightly saturated. Using our \hmol~fit model we carefully deblend the D\,I and \hmol~transitions. We use then the AOD technique to derive a lower limit to $N$(D\,I) using the deblended D\,I $\lambda$919.1 transition. This exercise yields log $N$(D\,I) $>$ 16.13, indicating that it is unlikely that $N$(D\,I) is overestimated.

We conclude then that the high D/H ratio along this sightline is unlikely due to either underestimation or overestimation of the H\,I or D\,I column densities, respectively.

 According to the variable deuterium depletion model proposed by \citet{Linsky2006}, sightlines that probe gas with high column densities should be biased by cold, not recently shocked gas, where deuterium would be depleted in grains. The high D/H toward HD\,41161 might indicate that this scenario occurs at $N$(H) higher than what was previously thought or that the clumping of low D/H values in the literature has another explanation. There is also the possibility that this sightline is dominated by gas which has been shocked and where the typical traces of a shock, such as highly ionized species (not detected along this sightline), have been erased. In this scenario the timescale to erase the traces of the shock would have to be smaller than that for D to deplete into the grains. The presence of cold \hmol~($T_{\rm 01}$ = 84 $\pm$ 7 K) along this sightline does not necessarily imply that the gas is cold and undisturbed since \hmol~is only a minor constituent of the sightline. Finally, we cannot rule out that local production of D\,I of the type proposed by \citet{1999ApJ...511..502M} or \citet{2003ApJ...597...48P} is responsible for all or part of the deuterium enhancement along this sightline.

To determine the iron depletion along this sightline, D(Fe) = Log (Fe/H)$_{\rm gas}$ - Log (Fe/H)$_{\odot}$, we use the solar abundance ratio Log (Fe/H)$_{\odot}$ = $-$4.55 $\pm$ 0.05 from \citet{2005ASPC..336...25A}. We find for this sightline, D(Fe) = $-$1.55 $\pm$ 0.11. In a manner similar to some other sightlines with high D/H ratios, the Fe/H ratio along this sightline does not fit the trend of high Fe/H versus high D/H, displayed for most of the cases shown in Figure 3 of \citet{Linsky2006}. There are however a few sightlines in Figure 3 of the paper mentioned above that seem to follow a trend different than most of the points, in that for the same Fe depletion a higher D/H ratio is found. The D/H and Fe/H ratios along the sightline to HD\,41161 seem to follow this trend as well. \citet{Linsky2006} discuss some of the reasons that could lead to this sort of behavior and argue that these sightlines could either contain a smaller percentage of carbon grains (thus fewer sites for deuterium to deplete) or could have had weak shocks pass by which would have evaporated deuterium from the grain mantles.

\subsubsection{D/H along the HD\,53975 sightline}

For the HD\,53975 sightline we derive D/H = (1.02 $\pm~^{0.23}_{0.20}$)$\times^{-5}$ in agreement with the weighted mean D/H = (0.86 $\pm$ 0.08)$\times10^{-5}$, quoted above. Although the D/H ratio along the HD\,41161 sightline indicates that D/H is variable for log $N$(H) $>$ 20.7, the D/H ratio along the HD\,53975 sightline follows the trend displayed by the other other five sightlines in the literature with log $N$(H) $>$ 20.7. Taking into account this ratio, along with the ratios for the five sightlines mentioned above, yields a weighted mean D/H = (0.89 $\pm$ 0.07)$\times10^{-5}$ (with $\chi^2_\nu$ = 0.4 for 5 degrees of freedom; the average of the upper and lower 1$\sigma$ uncertanties of individual ratios are used to determine weighted means in this work). The iron depletion along this sightline, D(Fe) = $-$1.75 $\pm$ 0.09, follows the trend of low Fe/H versus low D/H displayed in Figure 3 of \citet{Linsky2006}.

\subsection{O/H Ratios}

\begin{figure}
\begin{center}
\epsscale{0.75}
\rotatebox{90}{
\plotone{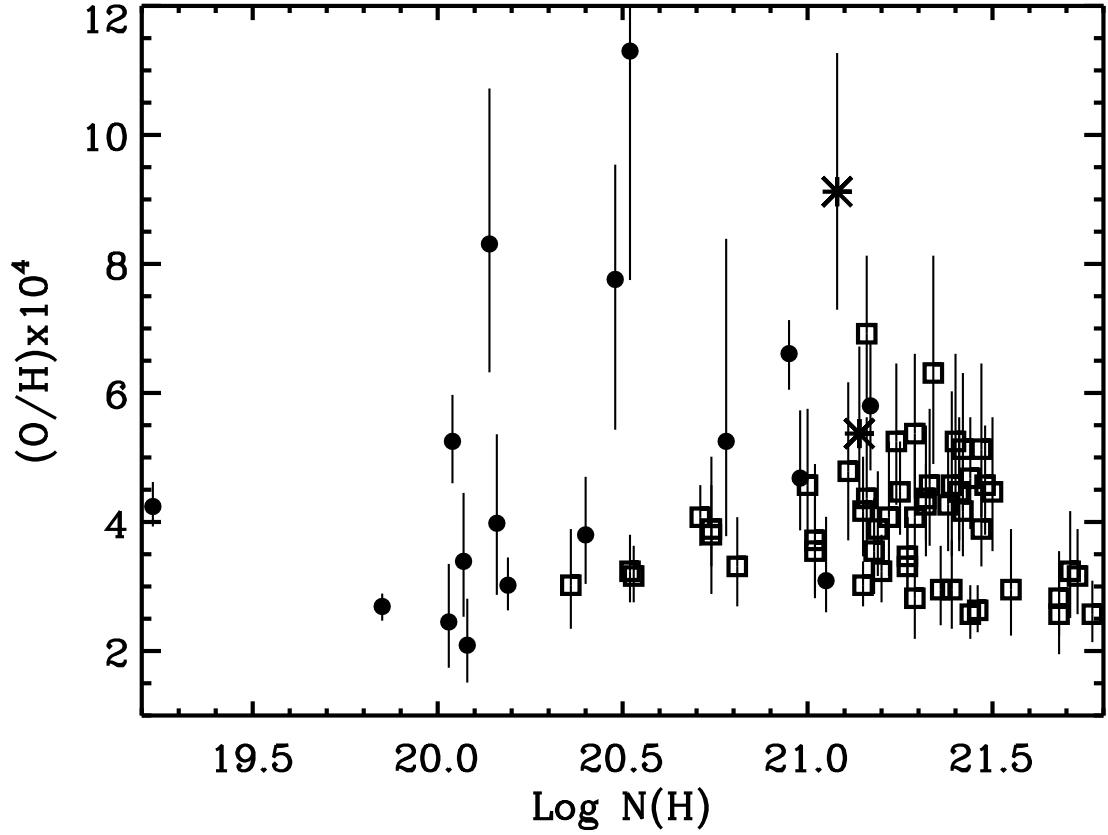}
}
\caption{O/H as a function of Log $N$(H). Filed circles correspond to {\it FUSE} based measurements from the literature \citep[see for e.g.][for a compilation of these values]{2006ApJ...642..283O}. The two new values derived here are represented by asterisks and open squares represent {\it HST} based measurements from \citet{2004ApJ...613.1037C}. \label{oiscatterplot}}
\end{center}
\end{figure}

The O/H ratio along the HD\,41161 sightline, O/H = (9.12 $\pm~^{2.15}_{1.83}$)$\times10^{-4}$, is more than 3$\sigma$ away from O/H = (3.43 $\pm$ 0.15)$\times10^{-4}$ derived for 13 lines of sight probing gas within 1500 pc (most within 500 pc), with 20.18 $\leq$ log $N$(H\,I) $\leq$ 21.28 by \citet{1998ApJ...493..222M} (updated $f$-value). For HD\,53975, O/H = (5.37 $\pm~^{1.35}_{1.14}$)$\times10^{-4}$, is $\sim$1.7$\sigma$ away from the \citet{1998ApJ...493..222M} O/H ratio. The {\it FUSE}-based O/H measurements present a larger scatter than those based on {\it HST} data \citep[see for example][]{2003ApJ...591.1000A,2004ApJ...613.1037C}. Figure \ref{oiscatterplot} presents a compilation of O/H measurements performed with {\it FUSE} \citep[from this work and the compilation by][asterisks and filled circles, respectively]{2006ApJ...642..283O} and {\it HST} data \citep[from][diamonds]{2004ApJ...613.1037C}. For the {\it FUSE} measurements we consider only sightlines that probe gas beyond the Local Bubble, i. e., log $N$(H) $>$ 19.2, since previous measurements by \citet{2005ApJ...625..232O} have shown that O/H = (3.45 $\pm$ 0.19)$\times10^{-4}$ in the LB. The assumption that all the {\it HST} O/H measurements can be described by a single mean leads to the weighted mean O/H = (3.51 $\pm$ 0.10)$\times10^{-4}$ with $\chi^2_{\nu}$ = 1.35 for 51 degrees of freedom. Assuming that the {\it FUSE} measurements can also be described by a single mean leads to the weighted mean O/H = (3.56 $\pm$ 0.13)$\times10^{-4}$ but with $\chi^2_{\nu}$ = 5.4 for 18 degrees of freedom. The scatter in the {\it FUSE} O/H measurements is more pronounced in the range 20.0 $\leq$ Log $N$(H) $\leq$ 20.9, where there are only a few {\it HST} measurements available. Hence, the scatter in the {\it FUSE} measurements might be due to there really being more scatter for lower column density sightlines, even if the O/H values for the LB show little scatter \citep{2005ApJ...625..232O}.

The larger \citep[than][]{1998ApJ...493..222M} sample of sightlines probed by the work of \citet{2004ApJ...613.1037C} contains some sightlines with O/H ratios $\sim$6$\times10^{-4}$. For example, HD\,91824 ($l$ = 285.70$^\circ$, $b$ = 0.07$^\circ$) and HD\,91983 ($l$ = 285.88$^\circ$, $b$ = 0.05$^\circ$) are both members of the Carina OB1 association (located at $d$ $\sim$ 2.5 kpc) and have O/H = (6.91 $\pm~^{1.21}_{0.92}$)$\times10^{-4}$ and (5.25 $\pm~^{1.21}_{0.98}$)$\times10^{-4}$, respectively \citep[][based on {\it HST} data]{2004ApJ...613.1037C}. HD\,90087 ($l$ = 285.16$^\circ$, $b$ = $-$2.13$^\circ$) is also part of the Carina OB1 association and is $\sim$100 pc away from the two other stars. For this sightline the {\it FUSE} derived O/H = (5.8 $\pm$ 1.0)$\times10^{-4}$ \citep{2005ApJ...635.1136H} is in good agreement with the O/H ratios derived for the other members of this association.

 In view of the \citet{2004ApJ...613.1037C} measurements, the O/H ratio along the HD\,53975 sightline does not seem to abnormally high, however the ratio along the line of sight to HD\,41161 does. $N$(O\,I) is determined using the only unsaturated O\,I transition in the {\it FUSE} bandpass, $\lambda$974, which is heavily blended with \hmol~$J$ = 2 and $J$ = 5. One might suspect that an erroneous $f$-value would lead to an inaccurate $N$(O\,I). However, previous studies by \citet{2005ApJ...635.1136H} indicate that $N$(O\,I) derived from this transition is consistent with $N$(O\,I) derived both from weaker and stronger O\,I transitions. Unknown blends with stellar lines are also the unlikely cause of a seemingly high $N$(O\,I) due to the high $v_{\rm sini}$ of HD\,41161. We suspect that unknown systematic effects when deriving $N$(O\,I) from the heavily blended $\lambda$974 O\,I transition might be responsible for at least part of the scatter in the {\it FUSE} based O/H measurements.

\subsection{N/H, O/N, and D/O Ratios}

We derive N/H = (8.32 $\pm~^{2.09}_{1.76}$)$\times10^{-5}$ and N/H = (5.07 $\pm~^{1.45}_{1.21}$)$\times10^{-5}$ for the HD\,41161 and HD\,53975 sightlines, respectively. The first ratio is consistent, within the uncertainties, with N/H = (7.5 $\pm$ 0.4)$\times10^{-5}$ derived by \citet{1997ApJ...490L.103M}. The N/H ratio for the HD\,53975 sightline is $\sim$1.7$\sigma$ away from the \citet{1997ApJ...490L.103M}.

We derive O/N = 10.96 $\pm~^{2.10}_{1.85}$ and O/N = 11.74 $\pm~^{3.36}_{2.80}$ along the HD\,41161 and HD\,53975 sightlines, respectively.  The O/N ratios for HD\,41161 and HD\,53975 are more than 3$\sigma$ and 2.7$\sigma$, respectively, away from O/N = 4.1 $\pm$ 0.3 derived using the values of O/H and N/H determined by \citet{1998ApJ...493..222M} and \citet{1997ApJ...490L.103M}. This is likely a consequence of the high O/H for the two sightlines, in addition to the low N/H for the HD\,53975 sightline.

We derive D/O = (2.29 $\pm~^{0.40}_{0.35}$)$\times10^{-2}$ and D/O = (1.91 $\pm~^{0.51}_{0.43}$)$\times10^{-2}$ along the HD\,41161 and HD\,53975 sightlines, respectively. Both of these ratios are consistent within the uncertainties with the weighted average D/O = (1.75 $\pm$ 0.18)$\times10^{-2}$ derived by \citet{2005ApJ...635.1136H} for the distant sightlines. The low dispersion of D/O at high $N$(H) has been used by \citet{2003ApJ...599..297H} and \citet{2006ASPC..348...47H} as an argument against a high D/H ratio being representative of the interstellar medium in the Solar neighborhood. However, as pointed out above, the large scatter in the {\it FUSE} derived O/H ratios seems to indicate that $N$(O\,I) might suffer from unknown systematic errors, which would affect the derived D/O ratios. If the high value of $N$(O\,I) measured along the HD\,41161 sightline is indeed too high due to systematic effects related to using the $\lambda$974 transition, then the D/O ratio along this sightline would be have to be revised upwards becoming inconsistent with the \citet{2005ApJ...635.1136H} D/O ratio for distant sightlines. This would be consistent with our claim, from D/H, that the deuterium abundance towards HD\,41161 is higher than that along the other log $N$(H) $>$ 20.7 sightlines.  

Finally, we would like to point out that there are other sightlines in which high D/H, low D/O, and high O/H ratios are also observed, similarly to the HD\,41161 sightline. These are Feige 110 \citep{2002ApJS..140...37F,2005ApJ...635.1136H}, PG\,0038$+$199 \citep{2005ApJ...625..210W}, and LSE\,44 \citep{2006ApJ...638..847F}. Discussions about the possible causes of these ratios are given in the papers quoted for each sightline.

\section{$f$-VALUES OF CHLORINE}
\label{cl_disc}

Chlorine has a unique chemistry, in which it can react exothermically with molecular hydrogen. Chlorine is ionized in regions where hydrogen is mostly in the atomic form, whereas in regions where there is a significant amount of molecular hydrogen, chlorine is mostly neutral with a small fraction in the HCl molecular form \citep{1974ApJ...190L..33J,1978ApJ...219..861J}. Cl\,I traces then the cold molecular components along the sight lines. Determining accurate Cl\,I column densities requires accurate oscillator strengths; however only a few transitions have experimentally determined $f$-values. The {\it FUSE} bandpass contains numerous transitions of Cl\,I, many of which are detected along the HD\,41161 sight line. In addition, the strong $\lambda$1347 Cl\,I transition is also detected in the {\it IUE} data for this star. \citet{1993ApJ...406..735S} determined the $f$-values of several Cl\,I transitions using beam-foil spectroscopy. While the $f$-value of Cl\,I $\lambda$1347 had been previously measured experimentally, $\lambda\lambda$ 1088, 1097 had only theoretically determined $f$-values. The work by \citet{1993ApJ...406..735S} presents then the first and only up to now, experimental measurement of the $f$-values of the $\lambda$1088, and $\lambda$1097 transitions, and the most accurate measurement of the $f$-value of $\lambda$1347.

Using two Cl\,I transitions which have experimentally determined $f$-values ($\lambda\lambda$ 1088, 1347) we place constraints on the $f$-values of other transitions whose currently adopted theoretical $f$-values \citep[see][and references therein]{2003ApJS..149..205M} are obviously incorrect. For two transitions that are optically thin, i. e., on the linear part of the COG, one can derive the relationship between their equivalent widths, $W_{\lambda _{\rm 1}}$/$W_{\lambda_ {\rm 2}}$ = ($f_{\rm 1}$$\times \lambda_{\rm 1}^{\rm 2}$)/($f_{\rm 2}$$\times \lambda_{\rm 2}^{\rm 2}$). We can then use the wavelengths and experimentally derived $f$-values (as well as their associated uncertainties) for the $\lambda$1088 and $\lambda$1347 Cl\,I transitions to derive the predicted ratio ($W_{\rm 1347}$/$W_{\rm 1088}$)$_{\rm pred}$. Table \ref{cl_ew} presents the atomic data and equivalent widths of the Cl\,I transitions detected along the HD\,41161 sight line as well as 1 $\sigma$ upper limits on the equivalent widths of non-detected transitions. The fourth column in the table lists the references for the $f$-values. Using the data on this table we derive ($W_{\rm 1347}$/$W_{\rm 1088}$)$_{\rm pred}$ = 2.90 $\pm$ 0.33, under the assumption that for the HD\,41161 sightline these two transitions lie in the linear part of the COG. Using the measured equivalent widths and their uncertainties from Table \ref{cl_ew} for these two transitions we derive ($W_{\rm 1347}$/$W_{\rm 1088}$)$_{\rm meas}$ = 2.35 $\pm$ 0.41. The agreement between the predicted and measured equivalent width ratios indicates that the linear COG assumption is a valid approximation for this sightline. Since Cl\,I traces the cooler components of a cloud, where hydrogen is mostly in molecular form, it is possible that there are more than one Cl\,I components, not resolved by {\it FUSE}. However, as shown by \citet{jenkins86}, there are cases in which the combined equivalent widths of a population of Gaussian-like interstellar absorption components exhibit a COG which closely mimics that of a single, pure Gaussian distribution in velocity.

Although it seems that the approach above is valid for this sight line we take a conservative viewpoint by assuming that Cl\,I $\lambda$1347 could be mildly saturated, but that Cl\,I $\lambda$1088 is not. We use then the equivalent width of this transition as well as its uncertainty to predict the $f$-values of the other weaker Cl\,I transitions quoted in Table \ref{cl_ew} using the laboratory measured $f_{\rm 1088}$ and its uncertainty. For the transition $x$ the predicted $f_x$ is then $f_x$ = $f_{\rm 1088}\times (W_x/W_{\rm 1088}) \times (\lambda_{\rm 1088}/\lambda_x)^2$. Blending of Cl\,I lines with stellar lines is not a concern for this sightline due to the high $v_{sin i}$ of this star (see Table \ref{star_properties}). In addition, careful modeling of the \hmol~and HD absorption along this sight line assures us that the Cl\,I lines studied here are free from blends with these molecular species. In particular, the $\lambda$1088 Cl\,I transition which falls near a CO transition ($\lambda$1087.868) is not affected by a blend with this molecular species as log $N$(CO) $<$ 13.5 along this sightline. The last column in Table \ref{cl_ew} presents the predicted $f$-values (and 1 $\sigma$ uncertainties), using the equation above, which can then be compared to the $f$-values from the literature quoted in the second column of this table.
 
Our results, presented in the last column of Table \ref{cl_ew}, indicate that some Cl\,I transitions need to have their $f$-values revised by large factors. For instance, the currently used $f$-value for $\lambda$1085 is underestimated by a factor of $\sim$38, while that for $\lambda$1031 is overestimated by a factor of $\sim$7. While the aim of this work is not to determine empirically the Cl\,I $f$-values in the {\it FUSE} bandpass, the set of revised $f$-values we present in Table \ref{cl_ew} should provide a better match to the observations and can be used in the future for comparison with new theoretical calculations. For a comprehensive study of the oscillator strenghts of many Cl\,I transitions, including several not covered here, we direct the reader to the work by Sonnentruker et al. (2006, in preparation).

Taking into account the uncertainties in the $f$-value of 1088 as well as on the measured equivalent width and assuming that this transition in on the linear COG we derive log $N$(Cl\,I) = 13.72 $\pm$ 0.04.

There is a weak absorption feature around  976 \AA~($W_{\lambda}$ = 19.5 $\pm$ 1.6 \AA) which we suspect to be from chlorine, as we have not been able to identify it with other lines of other atomic or molecular species. We use the \hmol~($J$ = 2) $\lambda$975.3507 and HD ($J$ = 0) $\lambda$975.5821 lines to derive the wavelength of this unknown transition by assuming that it traces the same absorption component as \hmol~and HD. The \fuse~wavelength solution is expected to be accurate to about 6 km s$^{-1}$ which at these wavelengths translates into $\sim$0.02 \AA. We derive then $\lambda$ = 976.17 \AA~for this unknown transition.

\section{SUMMARY}
\label{summary}

We have used {\it FUSE} data to determine the abundance of several atomic and molecular species along the sightlines to HD\,41161 and HD\.53975. Together with $N$(H\,I) from the literature, we derive D/H, N/H, O/H, and D/O ratios for these sightlines. We find that while the D/H ratio along the HD\,53975 sightline is consistent with five previous measurements at similar $N$(H), the ratio toward HD\,41161 is surprisingly high and presents the first evidence of D/H variations for sightlines with high $N$(H). The pattern of low D/H for sightlines with large $N$(H), pointed out by \citet{2003ApJ...599..297H} and emphasized by \citet{2004ApJ...609..838W} and \citet{2005ApJ...635.1136H} is based on a small number of sightlines. Taking into account that at this high $N$(H) there is a selection bias against high D/H measurements \citep{2006ApJ...642..283O} it is not surprising that only 1 out of 7 sightlines present high D/H ratios. As analyses of D/H along other high column density sightlines are being performed, one might expect that future measurements will also produce high D/H ratios, indicating that our understanding of the mechanisms responsible for the observed D/H pattern are still incomplete. Finally, the large number of Cl\,I transitions detected in the spectra of HD\,41161 allows to place significant constraints on the $f$-values of several transitions for which only theoretical values are available.

\acknowledgments

This work is based on data obtained for the Guaranteed Time Team by the NASA-CNES-CSA \fuse~mission operated by The Johns Hopkins University. Financial support to U. S. participants has been provided in part by NASA contract NAS5-32985 to Johns Hopkins University. Finacial support to G. H{\'e}brard has been provided by CNES. Based on observations made with the {\it International Ultraviolet Explorer}, obtained from the Data Archive at the Space Telescope Science Institute, which is operated by the Association of Universities for Research in Astronomy, Inc. under NASA contract NAS5-26555. Support for MAST for non-HST data is provided by the NASA Office of Space Science via grant NAG5-7584 and by other grants and contracts. The profile fitting procedure, Owens.f, used in this work was developed by M. Lemoine and the French \fuse~Team.

\bibliography{ms}
\bibliographystyle{apj}

\clearpage

\begin{deluxetable}{lccccccc}
\tablewidth{0pc}
\tablecaption{Stellar Properties \label{star_properties}}
\tablehead{ 
\colhead{Star}&\colhead{$d$}&\colhead{$l$}&\colhead{$b$}&\colhead{Sp. Type}&\colhead{$v_{\rm sin~i}$} &\colhead{$E_{\rm B-V}$}\\
\colhead{} &\colhead{(pc)}&\colhead{($^\circ$)}  &\colhead{($^\circ$)} &\colhead{} &  \colhead{(km s$^{-1}$)} &\colhead{}}
\startdata
HD\,41161 &	1253	& 164.97	& $+$12.89	& O8.0 V & 300 & 0.20	\\
HD\,53975& 	1318	& 225.68	& $-$2.32	& O7.5 V & 163 & 0.22	\\
\enddata
\tablecomments{All stellar properties are from \citet {1994ApJS...93..211D} except for $v_{\rm sin~i}$ which are from \citet{1997MNRAS.284..265H}. Distances are based on photometry and are probably accurate within $\sim$25\%.}
\end{deluxetable}

\begin{deluxetable}{lcccccc}
\tablewidth{0pc}
\tablecaption{Log of \fuse~observations \label{fuse_obs}}
\tablehead{ 
\colhead{Star} & \colhead{Program ID} & \colhead{Aperture}& \colhead{Mode} &\colhead{Time (s)} & \colhead{Date} &\colhead{CalFUSE}} 
\startdata
HD\,41161	& P1021001 & LWRS & HIST & 58 & 2001 Feb 20 & 3.0.7 \\	
\ldots		& P1021002\tablenotemark{a} & MDRS & HIST & 6520 & 2003 Sept 25 & 3.0.8 \\	
HD\,53975	& P3032301\tablenotemark{a} & MDRS & HIST & 482 & 2004 Feb 04 & 3.0.7 \\
\enddata
\tablenotetext{a}{Only SiC data obtained during these observations.}
\end{deluxetable}

\begin{deluxetable}{lcccc}
\tablewidth{0pc}
\tablecaption{Atomic data and equivalent widths (m\AA) for the lines used in the analyses\tablenotemark{a} \label{atomicdata}}
\tablehead{ 
\colhead{Species} & \colhead{Wavelength (\AA)} & \colhead{Log $f\lambda$} & \colhead{HD\,41161} &\colhead{HD\,53975}}
\startdata
%
D\,I	& 916.1785	&$-$0.28& (A, P)	& (A, P)\\
\ldots	& 916.9311	&$-$0.18 & \ldots	& (P) \\
\ldots	& 917.8797	&$-$0.07& (P)	& \ldots\\
\ldots	& 919.1013	& 0.04  & \ldots	& (P)\\
\ldots	& 920.713	& 0.17	& \ldots	& (A, P)\\
C\,I	& 945.1910	& 2.16	& 89.59 $\pm$ 2.55 (A)	& 51.37 $\pm$ 4.30 (A) \\
C\,III	& 977.0200	& 2.87	& 222.90 $\pm$ 6.22 (A)	& 311.13 $\pm$ 17.61 (A)\\
N\,I	& 951.0792	&$-$0.79& 67.53 $\pm$ 2.42 (C, P)	& 49.96 $\pm$ 4.01 (C, P) \\
\ldots	& 951.2948	&$-$1.66& 18.52 $\pm$ 1.95 (A, C, P)	& 24.68 $\pm$ 3.76 (C, P)\\
\ldots  & 953.4152	& 1.09	& 126.05 $\pm$ 2.37 (C)	& 144.00 $\pm$ 3.45 (C)\\
\ldots	& 953.6549	& 1.37	& 132.32 $\pm$ 2.60 (C)	& 147.30 $\pm$ 3.41 (C)\\
\ldots	& 959.4937	&$-$1.30& 28.89 $\pm$ 1.26 (C, P)	& 22.55 $\pm$ 2.39 (A, C, P) \\
\ldots	& 960.2014	&$-$1.95& \ldots	& (P)	\\
\ldots	& 963.9903	& 1.08	& \ldots		& 154.30 $\pm$ 5.03 (C)\\
\ldots	& 964.6256	& 0.88	& \ldots 		& 128.36 $\pm$ 8.29 (C)\\
\ldots  & 1134.1653	& 1.22	& 158.26 $\pm$ 4.92 (C)	& \ldots \\
\ldots  & 1134.4149	& 1.51  & 162.92 $\pm$ 5.15 (C)	& \ldots \\
\ldots  & 1134.9803	& 1.67  & 180.55 $\pm$ 6.35 (C)	& \ldots \\
N\,II	& 1083.994	& 2.10	& 170.00 $\pm$ 4.94 (A)	& 244.55 $\pm$ 9.95 (A)\\
O\,I	& 974.070	& $-$1.82 & (P)	& (P) \\
O\,VI	& 1031.9261	& 2.13	&\ldots	& 91.16 $\pm$ 12.84 (P)\\
\ldots  & 1037.6167	& 1.83	&\ldots	& 42.92 $\pm$ 7.22 (A, P)\\
Mg\,II	& 946.7033	& $-$0.17 & \ldots & (P) \\
\ldots	& 946.7694	& $-$0.48 & \ldots & (P) \\	

P\,II	& 961.0412	& 2.53	& 74.90 $\pm$ 2.23 (A) & 88.23 $\pm$ 4.74 (A)\\
S\,III	& 1012.4950	& 1.65	& \ldots& 86.38 $\pm$ 10.13 (P)\\
Cl\,I	& 1088.0589	& 1.95	& 44.55 $\pm$ 1.97 (A)	& 14.16 $\pm$ 3.46 (A) \\
Cl\,II	& 1071.0358	& 1.21	& 16.16 $\pm$0.96 (A) & \ldots\\
Ar\,I	& 1066.6599	& 1.86	& 114.17 $\pm$ 3.03 (A)	& 158.12 $\pm$ 7.86 (A)\\
Ar\,II	&  919.7810	& 0.91	& \ldots	& (P)	\\
Fe\,II	& 926.8969	& 0.72	& \ldots			& 19.76 $\pm$ 2.06 (C, P)\\	
\ldots	& 1055.2617	& 0.90	& 46.34 $\pm$ 1.31 (C, P)	& 46.33 $\pm$ 2.44 (A, C, P) \\
\ldots	& 1062.1517	& 0.49	& 17.59 $\pm$ 2.17 (C, P)	& \ldots \\ 
\ldots	& 1063.9718	& 0.60	& 20.41 $\pm$ 2.19 (C)	& 19.18 $\pm$ 3.70 (A, C) \\
\ldots	& 1081.8748	& 1.13	& 88.26 $\pm$ 2.36 (C)	& 61.79 $\pm$ 3.59 (C) \\
\ldots	& 1083.4204	& 0.48	& 27.75 $\pm$ 2.22 (C, P) & \ldots \\
\ldots	& 1096.8770	& 1.55	& 93.16 $\pm$ 3.03 (C)	& \ldots \\
\ldots	& 1112.0480	& 0.84	& 34.34 $\pm$ 3.78 (C)	& \ldots \\
\ldots	& 1121.9749	& 1.36	& 83.15 $\pm$ 5.79 (C)	& \ldots \\
\ldots	& 1125.448	& 1.26	& 84.07 $\pm$ 4.35 (C)	& \ldots \\
\ldots	& 1127.0984	& 0.50	& 25.86 $\pm$ 5.16 (A, C)	& \ldots \\
\ldots	& 1133.6654	& 0.80	& 50.57 $\pm$ 3.73 (A, C, P)	& \ldots \\
\ldots	& 1142.3656	& 0.68	& 45.39 $\pm$ 4.79 (C, P)	& \ldots \\
\ldots	& 1143.2260	& 1.31	& 80.79 $\pm$ 4.64 (C,P)	& \ldots \\
\ldots	& 1144.938	& 2.08	& 122.30 $\pm$ 5.36 (C)	& \ldots \\
\enddata
\tablenotetext{a}{A, C, and P indicate which methods were used with each transition and stand for apparent optical depth, curve-of-growth, and profile fitting, respectively.}
\end{deluxetable}

\begin{deluxetable}{lcccc}
\tablewidth{0pc}
\tablecaption{Atomic Column Densities (Log) for HD\,41161\label{NHD41161}}
\tablehead{ 
\colhead{Species} & \colhead{AOD} & \colhead{COG}& \colhead{PF} & \colhead{Adopted}} 
\startdata
H\,I	& \ldots & \ldots & \ldots & 21.00 $\pm$ 0.09\tablenotemark{a}\\
D\,I	& 16.37 $\pm$ 0.04 ($\lambda$916.18)	& \ldots	& 16.43 $\pm$ 0.03	& 16.40 $\pm$ 0.05 \\
N\,I	& 17.06 $\pm$ 0.04 ($\lambda$951.29)	& 17.00 $\pm~^{0.04}_{0.03}$ & 16.94 $\pm$ 0.01 & 17.00 $\pm$ 0.06\\
O\,I	& \ldots	& \ldots	& 18.04 $\pm$ 0.05 & 18.04 $\pm$ 0.05 \\
Fe\,II	& 14.96 $\pm~^{0.08}_{0.10}$ ($\lambda$1127)	& 15.02 $\pm$ 0.06	& 14.94 $\pm$ 0.03 & 14.98 $\pm$ 0.06	\\
Cl\,I	& 13.72 $\pm$ 0.04 ($\lambda$1088)	& \ldots	& \ldots	& 13.72 $\pm$ 0.04	\\
\hline
C\,I	& $\ge$	14.13 ($\lambda$945)	& \ldots	& \ldots 	& $\ge$	14.13 \\
C\,III	& $\ge$ 14.04 ($\lambda$977) & \ldots	& \ldots	& $\ge$ 14.04 \\
N\,II	& $\ge$ 14.52 ($\lambda$1083) & \ldots	& \ldots	& $\ge$ 14.52 \\
P\,II	& $\ge$ 13.63 ($\lambda$961) & \ldots	& \ldots	& $\ge$ 13.63 \\
Cl\,II	& $\ge$ 14.02 ($\lambda$1071) &\ldots	& \ldots	& $\ge$ 14.02 \\
Ar\,I	& $\ge$ 14.52 ($\lambda$1066) & \ldots	& \ldots	& $\ge$ 14.52 \\
\enddata
\tablenotetext{a}{See $\S$\ref{hd41161los}.}
\tablecomments{All uncertainties are 1$\sigma$.}
\end{deluxetable}


\begin{deluxetable}{lcc}
\tablewidth{0pc}
\tablecaption{Molecular Column Densities (Log) \label{h2col}}
\tablehead{ 
\colhead{Species} & \colhead{HD\,41161} & \colhead{HD\,53975}} 
\startdata
HD ($J$ = 0)	& 14.57 $\pm$ 0.08 & 13.42 $\pm$ 0.10 \\
\hmol~($J$~= 0)	& 19.63 $\pm$ 0.06 &  18.79 $\pm$ 0.05\\
\ldots ($J$~= 1)& 19.72 $\pm$ 0.06 & 18.95 $\pm$ 0.05\\
\ldots ($J$~= 2)& 17.97 $\pm$ 0.10 & 17.8: \\
\ldots ($J$~= 3)& 17.40 $\pm$ 0.10 & 17.7:\\
\ldots ($J$~= 4)& 14.95 $\pm$ 0.08 & 14.33 $\pm$ 0.05\\
\ldots ($J$~= 5)& 14.34 $\pm$ 0.05 & 13.80 $\pm$ 0.05 \\
\hline
$N$(\hmol)	& 19.98 $\pm~^{0.08}_{0.09}$ & 19.18 $\pm$ 0.04\\
$N$(H)		& 21.08 $\pm$ 0.08 & 21.14 $\pm$ 0.06 \\
\enddata
\tablecomments{All uncertainties are 1$\sigma$.}
\end{deluxetable}

\begin{deluxetable}{lcccc}
\tablewidth{0pc}
\tablecaption{Atomic Column Densities (Log) for HD\,53975 \label{NHD53975}}
\tablehead{ 
\colhead{Species} & \colhead{AOD} & \colhead{COG}& \colhead{PF} & \colhead{Adopted}} 
\startdata
H\,I	& \ldots & \ldots & \ldots & 21.13 $\pm$ 0.06\tablenotemark{a}\\
D\,I	& 16.11 $\pm$ 0.03 ($\lambda$920.7) & \ldots & 16.20 $\pm$ 0.02 & 16.15 $\pm$ 0.07 \\
N\,I  & 16.73 $\pm$ 0.05 ($\lambda$959.49) & 16.89 $\pm~^{0.10}_{0.07}$ & 16.84 $\pm$ 0.05 & 16.80 $\pm$ 0.08\\
O\,I  & \ldots	& \ldots & 17.87 $\pm$ 0.08 & 17.87 $\pm$ 0.08 \\
O\,VI	& 13.97 $\pm$ 0.07 ($\lambda$1031)	& \ldots & 13.96 $\pm$ 0.03 & 13.96 $\pm$ 0.05  \\
Fe\,II & 14.80 $\pm~^{0.07}_{0.08}$ ($\lambda$1063.9) & 14.87 $\pm~^{0.08}_{0.11}$ & 14.84 $\pm~^{0.02}_{0.03}$ & 14.84 $\pm$ 0.05 \\ 
Mg\,II	& \ldots	& \ldots 	& 16.10 $\pm$ 0.05	& 16.10 $\pm$ 0.05 \\
Cl\,I	& 13.53 $\pm$ 0.10 ($\lambda$1088)	& \ldots	& \ldots	& 13.53 $\pm$ 0.10\\
\hline
C\,I	& $\ge$ 13.73 ($\lambda$945)	& \ldots & \ldots & $\ge$ 13.73	\\
C\,III	& $\ge$ 14.22 ($\lambda$977)	& \ldots & \ldots & $\ge$ 14.22	\\
N\,II	& $\ge$ 14.73 ($\lambda$1083)	& \ldots & \ldots & $\ge$ 14.73	\\
P\,II 	& $\ge$ 13.62 ($\lambda$961)	& \ldots & \ldots & $\ge$ 13.62	\\
S\,III 	& $\ge$ 14.54 ($\lambda$1012)	& \ldots & \ldots & $\ge$ 14.54	\\
Ar\,I	& $\ge$ 14.67 ($\lambda$1066)	& \ldots & \ldots & $\ge$ 14.67	\\
\enddata
\tablenotetext{a}{See $\S$ \ref{hd53975los}.}
\tablecomments{All uncertainties are 1$\sigma$.}
\end{deluxetable}

\begin{deluxetable}{lcc}
\tablewidth{0pc}
\tablecaption{Ratios of Column Densities \label{ratios}}
\tablehead{ 
\colhead{Ratio} & \colhead{HD\,41161} & \colhead{HD\,53975}} 
\startdata
(D/H)$\times10^{5}$ & 2.14 $\pm~^{0.51}_{0.43}$	& 1.02 $\pm~^{0.23}_{0.20}$	\\
(N/H)$\times10^{5}$ & 8.32 $\pm~^{2.09}_{1.76}$	& 5.07 $\pm~^{1.45}_{1.21}$	\\
(O/H)$\times10^{4}$ & 9.12 $\pm~^{2.15}_{1.83}$	& 5.37 $\pm~^{1.35}_{1.14}$	\\ 
(Fe/H)$\times10^{6}$ & 0.79 $\pm~^{0.20}_{0.17}$ & 0.50 $\pm~^{0.10}_{0.08}$	\\
(D/O)$\times10^{2}$ & 2.29 $\pm~^{0.40}_{0.35}$	& 1.91 $\pm~^{0.51}_{0.43}$	\\ 
O/N & 10.96 $\pm~~^{2.10}_{1.85}$ & 11.74 $\pm~^{3.36}_{2.80}$ \\
(HD/\hmol)$\times10^{6}$	& 3.89 $\pm~^{1.11}_{0.93}$ &	1.74 $\pm~^{0.48}_{0.39}$ \\
$f_{\rm H_2}$	(\%)	& 16.0 $\pm~^{4.0}_{3.4}$ &	2.2 $\pm~^{0.4}_{0.3}$	\\
$\langle n_{\rm H}\rangle$ (cm$^{-3}$)	& 0.31 	& 0.34		\\	
\enddata
\tablecomments{All uncertainties are 1$\sigma$ and D/H = [$N$(D\,I) $+$ $N$(HD)]/[$N$(H\,I) $+$ 2$\times$$N$(\hmol) $+$ $N$(HD)]. }
\end{deluxetable}


\begin{deluxetable}{lcccc}
\tablewidth{0pc}
\tablecaption{Equivalent Widths and Atomic Data for Cl\,I along the HD\,41161 Sightline \label{cl_ew}}
\tablehead{ 
\colhead{Wavelength (\AA)} & \colhead{$f$} & \colhead{Ref.} & \colhead{$W_{\lambda}$ (m\AA)}  &\colhead{$f_{\rm pred.}$\tablenotemark{a}}} 
\startdata
969.9195	& 1.73$\times10^{-2}$		& 1	& 4.38 $\pm$ 0.93	& (1.0 $\pm$ 0.2)$\times10^{-2}$ \\	
984.2865\tablenotemark{b}& 2.44$\times10^{-2}$	& 1, 4	& 13.18 $\pm$ 1.95	& (2.9 $\pm$ 0.5)$\times10^{-2}$ \\
1004.6776	& 1.58$\times10^{-1}$		& 1, 4	& 28.63 $\pm$ 1.39	& (6.1 $\pm$ 0.7)$\times10^{-2}$ \\
1022.0478	& 1.22$\times10^{-2}$		& 1	& 9.35 $\pm$ 1.24	& (1.9 $\pm$ 0.3)$\times10^{-2}$	 \\
1027.3386	& 6.58$\times10^{-3}$		& 1	& 12.08 $\pm$ 2.97	& (2.5 $\pm$ 0.7)$\times10^{-2}$	 \\
1031.5070	& 1.51$\times10^{-1}$		& 1, 5	& 10.23 $\pm$ 3.42	& (2.1 $\pm$ 0.7)$\times10^{-2}$ \\
1085.3035	& 5.81 $\times10^{-4}$		& 2, 5	& 12.32 $\pm$ 2.16	& (2.2 $\pm$ 0.5)$\times10^{-2}$ \\
1088.0589	& (8.1 $\pm$ 0.7)$\times10^{-2}$& 3, 5	& 44.55 $\pm$ 1.97  	& \ldots	 \\
1094.7686	& 1.66$\times10^{-2}$		& 2, 5	& 40.21 $\pm$ 3.04	& (7.2 $\pm$ 0.9)$\times10^{-2}$ \\
1347.2396	& (1.53 $\pm$ 0.11)$\times10^{-1}$& 3, 5& 104.9 $\pm$ 17.4	  & \ldots	\\
\hline
978.5932	& 1.85$\times10^{-1}$ & 1, 5	&	$<$4.5		& $<$1.0$\times10^{-2}$ \\
1043.9860	& 1.81$\times10^{-2}$ & 1, 5	&	$<$2.7		& $<$5.3$\times10^{-3}$ \\
1079.8821	& 6.98$\times10{-3}$  &	2, 5	& 	$<$4.1		& $<$7.5$\times10^{-3}$ \\
1099.5230	& 1.14$\times10^{-2}$	& 2, 5	&	$<$10.0 	& $<$1.8$\times10^{-2}$	\\
1101.9361	& 1.13$\times10^{-2}$	& 2, 5	&	$<$5.1 	 	& $<$9.1$\times10^{-3}$	\\
\enddata
\tablenotetext{a}{See $\S$ \ref{cl_disc} for a discussion of how $f_{\rm pred.}$ is calculated.}
\tablenotetext{b}{Possibly blended with the $\lambda$984.323 transition.}
\tablecomments{For the last six transitions we quote 1$\sigma$ upper limits on the equivalent widths.}
\tablerefs{References 1--3 refer to the original papers. References 4 and 5 are compilations of atomic data. (1) \citet{1975SAOSR.362.....K}, (2) \citet{1994A&A...287..290B}, (3) \citet{1993ApJ...406..735S}, (4) \citet{1991ApJS...77..119M}, (5) \citet{2003ApJS..149..205M}.}
\end{deluxetable}
\clearpage

\end{document}